\documentclass[iop,twocolumn]{emulateapj}   
\bibpunct[; ]{(}{)}{;}{a}{}{;}              
\usepackage{apjfonts}                       

\usepackage{graphicx}
\usepackage{natbib}
\usepackage{amsmath}
\usepackage{upgreek}

\newcommand{\Msolar}{M${_\odot}$}

\shorttitle{Extinction in the LMC bar}
\shortauthors{De Marchi, Panagia, Milone}

\received{10 July 2021}
\accepted{23 September 2021}

\begin{document}

\title{Extinction in the Large Magellanic Cloud bar around NGC\,1854,
NGC\,1856, and NGC\,1858\,\altaffiliation{Based on
observations with the NASA/ESA {\it Hubble Space Telescope}, obtained at the
Space Telescope Science Institute, which is operated by AURA, Inc., under
NASA contract NAS5-26555}}

\author{Guido De Marchi}
\affiliation{European Space Research and Technology Centre,
Keplerlaan 1, 2200 AG Noordwijk, Netherlands; gdemarchi@esa.int}
\author{Nino Panagia}
\affiliation{Space Telescope Science Institute, 3700 San Martin
Drive, Baltimore MD 21218, USA; panagia@stsci.edu}
\affiliation{Supernova Limited, OYV \#131, Northsound Rd., Virgin
Gorda, British Virgin Islands}
\author{Antonino P. Milone}
\affiliation{Dipartimento di Fisica e Astronomia,
Univ. di Padova, Vicolo dell'Osservatorio 3, Padova I-35122, Italy;
antonino.milone@unipd.it}
\affiliation{Istituto Nazionale di Astrofisica -- Osservatorio
Astronomico di Padova, Vicolo dell'Osservatorio 5, Padova I-35122,
Italy}

\begin{abstract}  

We report on the extinction properties in the fields { around} the
clusters NGC\,1854, NGC\,1856, and NGC\,1858 in the bar of the Large
Magellanic Cloud (LMC). The colour--magnitude diagrams of the stars in
all these regions show an elongated red giant clump (RC) that reveals a
variable amount of extinction across these fields, ranging from
$A_V\simeq 0.2$ to $A_V\simeq 1.9$, including Galactic foreground
extinction. The extinction properties nonetheless are remarkably
uniform. The slope of the reddening vectors measured in the  ($V-I$,
$V$) and ($B-I$, $B$) colour--magnitude planes is fully in line with the
$A_V/E(B-V)\simeq 5.5$ value found in the outskirts of 30\,Dor. This
indicates the presence of an additional grey extinction component in the
optical requiring big grains to be about twice as abundant as in the
diffuse Galactic interstellar medium (ISM). Areas of higher extinction
appear to be systematically associated with regions of more intense star
formation, as measured by the larger number of stars more massive than
8\,\Msolar, thus making injection of big grains into the ISM by SNII
explosion the likely mechanism at the origin of the observed grey
extinction component.

\end{abstract}


\section{Introduction}

An investigation of the extinction properties towards a number of
regions of recent star formation in the Large Magellanic Cloud (De
Marchi \& Panagia 2014, 2019; De Marchi et al. 2014, 2016, 2020) has so
far revealed a consistent picture: shortwards of $\sim 1\,\mu$m,  the
extinction curve is systematically flatter (in logarithmic units) than
in the diffuse interstellar medium (ISM). This points to the presence of
a grey component due to a larger proportion of big grains. In the
regions studied so far (30 Dor, NGC 2060, NGC 1938), the total
extinction is not only uneven, but also rather large for the LMC, with
$A_V\la2$.

We have undertaken a systematic study of all the LMC star-forming
regions for which high-quality photometry with the {\em Hubble Space
Telescope} (HST) is available, in an attempt to understand whether and
how the extinction properties depend on the actual amount of extinction
present in these fields. In the Milky Way, regions of heavy extinction
along the Galactic plane have long been known to show a ratio of
total-to-selective extinction  $R_V \equiv A_V/E(B-V)$ systematically
larger than the value of $\sim 3.1$ typical of the diffuse ISM. Examples
include the Orion Nebula, NGC\,2244, I Ara, IC\,2851 (see, e.g.,
Sharpless 1952; Sharpless 1962; Johnson 1965; Turner 1973; Herbst 1975).
In this work we consider three additional LMC clusters located in the
bar of the galaxy, which are subject to intermediate values of $A_V$, in
order to explore whether the extinction properties and the value of
$R_V$ depend indeed on the amount of dust in these fields.

The first is the cluster NGC\,1858. The only study so far of the stellar
populations in this cluster based on digital photometry is that of
Vallenari et al. (1994). Adopting $E(B-V)=0.15\pm0.05$ (from Caplan \&
Deharveng 1984), these authors show that the most likely age of the main
cluster population is $8-10$\,Myr, although the presence of a protostar
in the field (Epchtein et al. 1984) suggests that some star formation
might still be active.

The second is the cluster NGC\,1854 (also called NGC\,1855). In a study
based on photoelectric photometry, Connolly \& Tifft (1977) showed that
the cluster has a tight main sequence (MS) and derived an 
age of $25 \pm 15$ Myr, with a colour excess $E(B-V)$ about $0.1$ mag
redder than the neighbouring field. The age determination is fully
consistent with later works based on photographic photometry by Hodge
(1983), suggesting $30\pm10$\,Myr, and by Alcaino \& Liller (1987), who
derived $25\pm6$\,Myr. No results based on digital photometry exist in
the literature for this cluster.

The third cluster is NGC\,1856, whose stellar population and extended 
MS turn-off in the colour-magnitude diagram (CMD) have been extensively 
studied in the past decade. Works by Bastian \& Silva--Villa (2013),
D'Antona et al. (2015), Correnti et al. (2015), and Milone et al. (2015)
concur in assigning to it an age of $\sim 300$\,Myr. Correnti et al.
(2015) and Milone et al. (2015) address the presence of uneven
extinction across the cluster. They assume that the extinction
properties, namely the extinction law and direction of the reddening
vector, are those of the ``standard'' MW extinction curve. In this work
we will not cover the NGC\,1856 cluster itself, but rather four regions
adjacent to it, at a projected distance of $\sim 5^\prime$ or $\sim
75$\,pc, specifically to probe an area still relatively close to
NGC\,1854 and NGC\,1858 (about 200 pc to the south) but not affected by
recent massive star formation episodes.

\begin{figure*}
\centering
\resizebox{\hsize}{!}{\includegraphics[height=13.1cm]{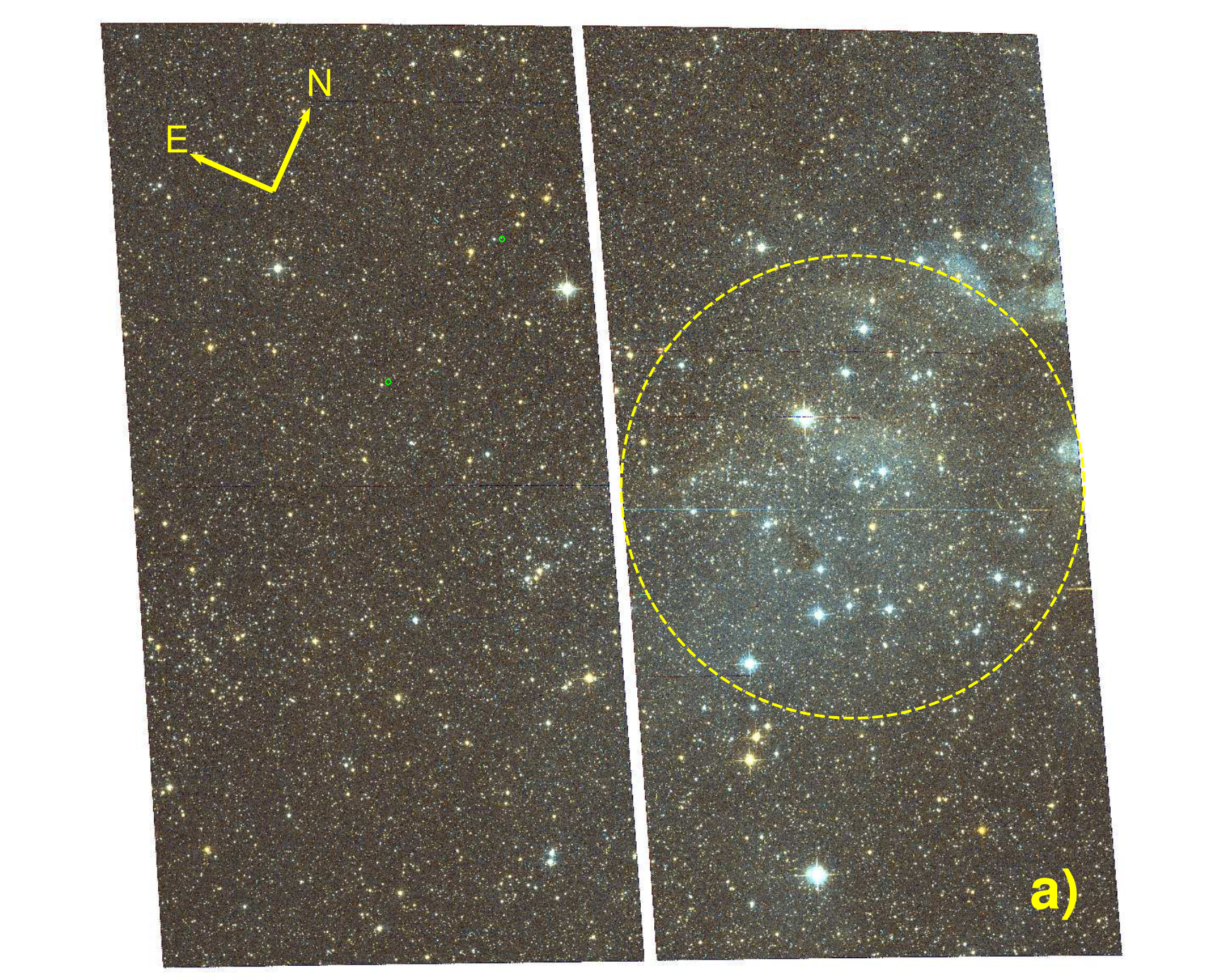}
                      \includegraphics[height=13cm]{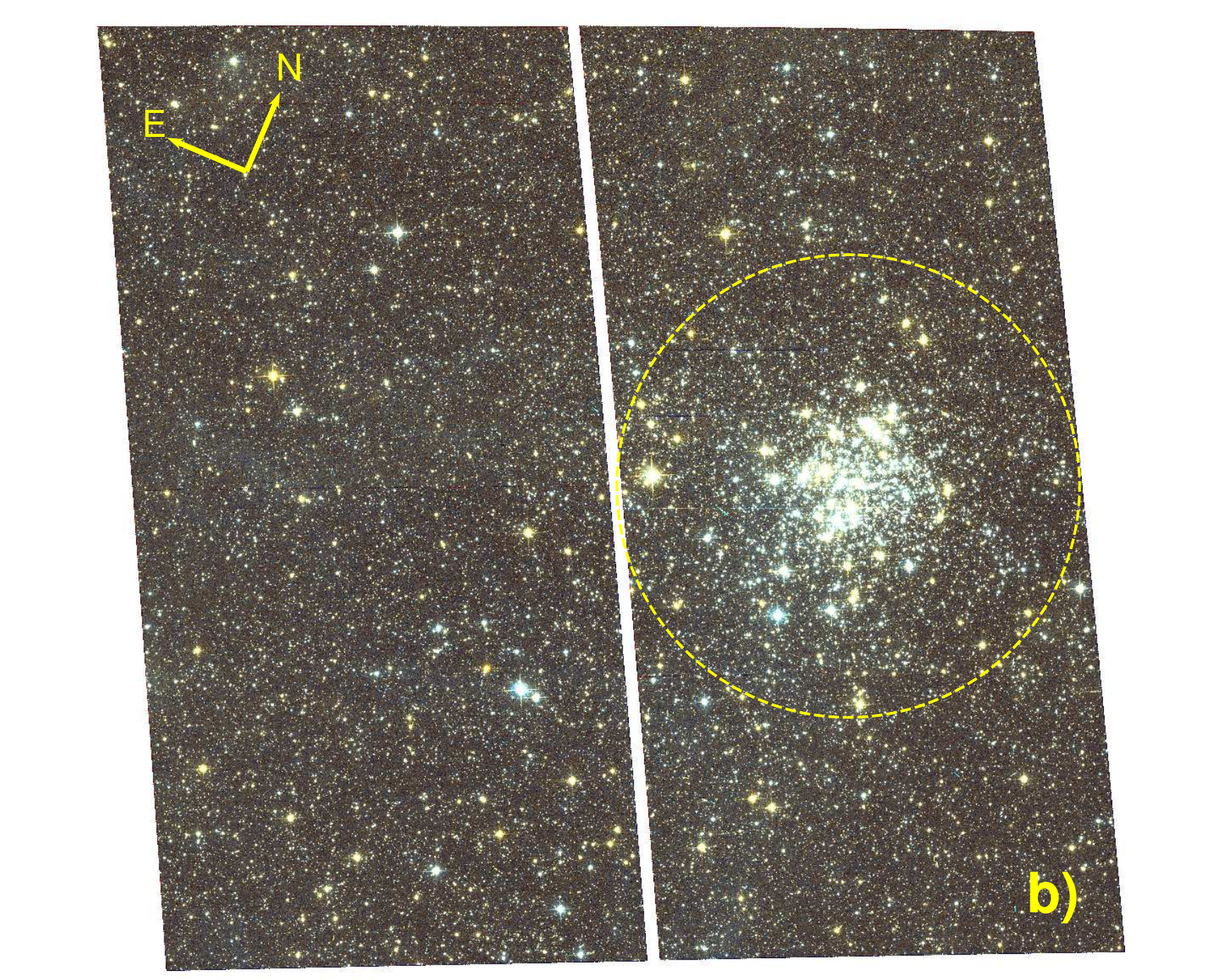}}
\resizebox{\hsize}{!}{\includegraphics[width=14cm]{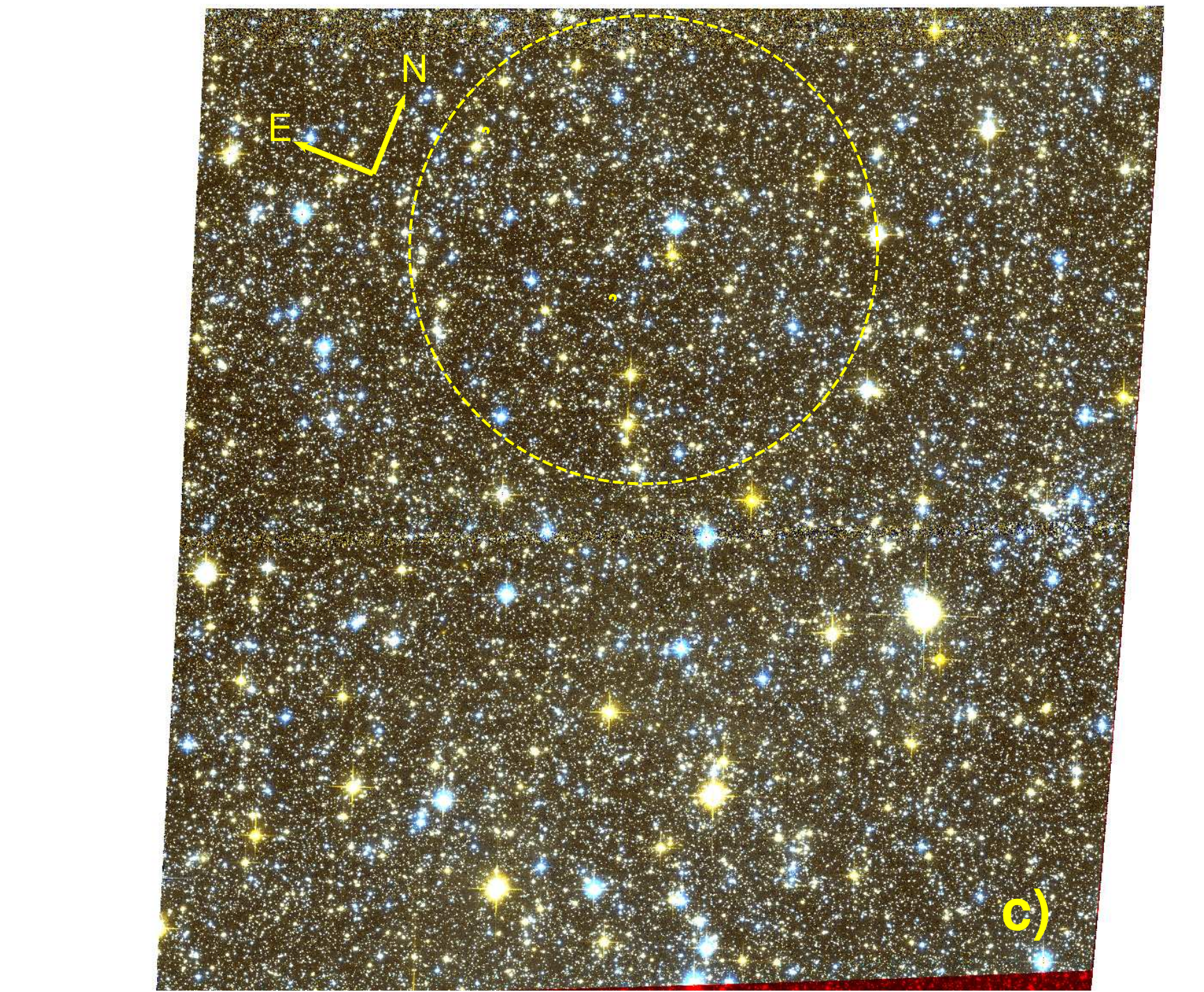}
  \includegraphics[trim=-1.5cm 0cm 1cm 0cm,clip,width=15.4cm]{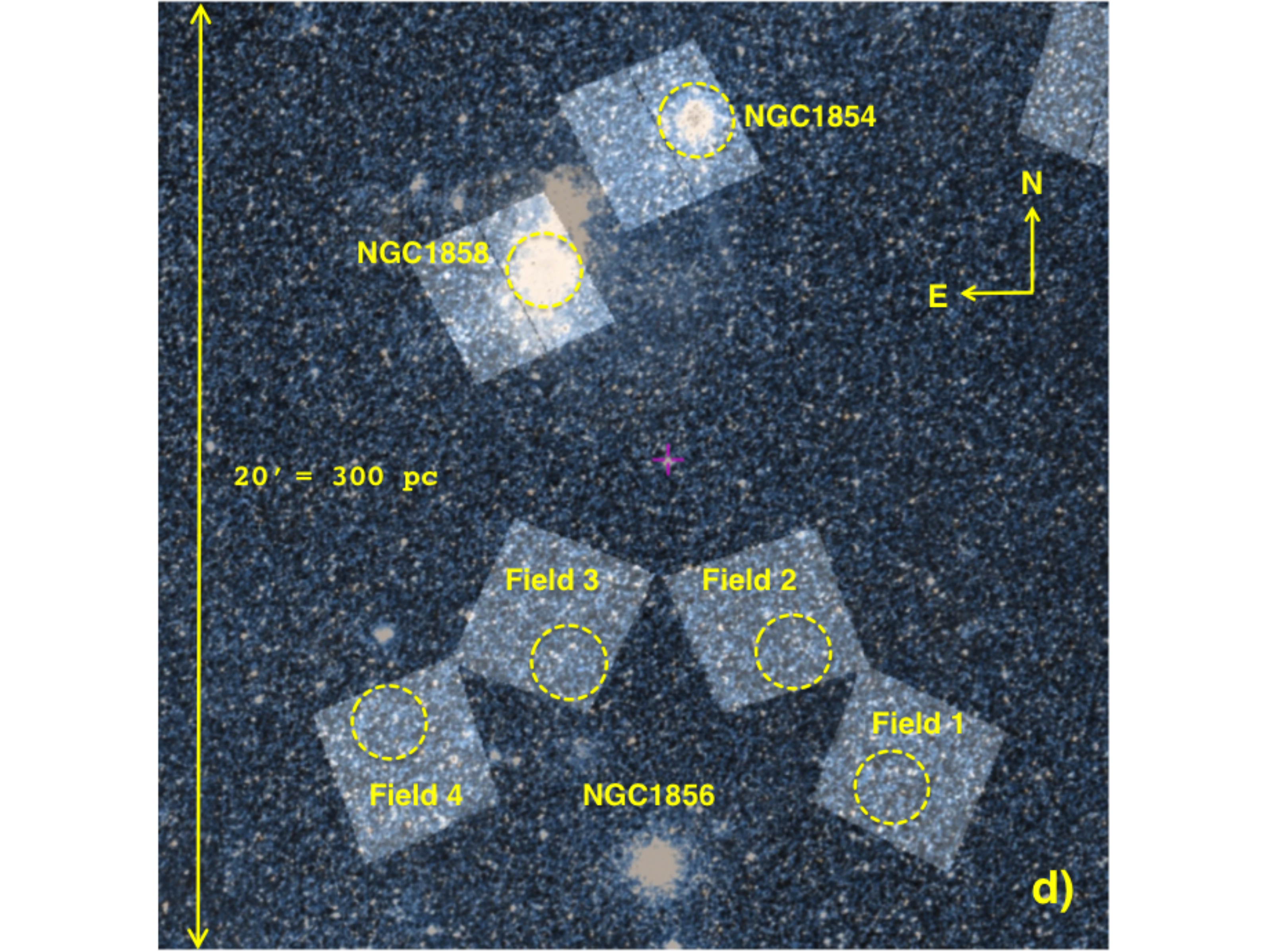}}
\caption{Colour composite images of the  regions studied in this work.
Panel a) and b), corresponding respectively to fields around NGC\,1858
and NGC\,1854, were obtained by combining the exposures in the $V$ and
$I$ bands. Panel c) corresponds to Field 4, located to the north-east of
NGC\,1856, and was obtained from the combination of the $B$ and $I$
bands. The three fields span  approximately $205\arcsec$ or $\sim
50$\,pc on a side. Panel d) shows the projected distribution of the
fields on the plane of the sky, in a region of about $20\arcmin$ or
$\sim 300$\,pc on a side. The yellow circles, with a radius of
$50\arcsec$ of $12.5$ pc, indicate the areas corresponding to the CMDs
shown in Figures\,\ref{fig2} -- \ref{fig4}.} 
\label{fig1}
\end{figure*}

In this paper, we investigate the origin of the patchy extinction
causing differential reddening across the fields around NGC\,1854,
NGC\,1856, and NGC\,1858 and show that the extinction properties, and
the grain size distribution that they imply, are consistent with those
measured in a field North of NGC\,2060. The structure of the paper is as
follows. In Section\,2 we present the observations and their analysis.
In Section\,3 we discuss the different populations present in the
fields, while Section\,4 is devoted to the extinction properties, which
we compare with those of other regions in the LMC. A discussion and the
conclusions follow, respectively, in Sections\,5 and 6.

\vspace*{0.5cm}
\section{Observations and data analysis}

The { fields around the} clusters NGC\,1854, NGC\,1856, and NGC\,1858
were observed with the {\em Wide Field Channel} (WFC) of the {\em
Advanced Camera for Surveys} (ACS) on board the HST. Colour composite
images of the  regions studied in this work are shown in
Figure\,\ref{fig1}, while details on the filters, exposure times, and
dates of the observations are contained in Table\,\ref{tab1}. 

The effective point spread function (ePSF) fitting procedure developed
by Anderson et al. (2008) was used for the astrometric and  photometric
analysis of the images. The stellar positions derived in this way were
further corrected for geometric distortion by using the solution by
Anderson \& King (2006). The instrumental magnitudes were calibrated in
the VEGAMAG reference system following Anderson et al. (2008), with the
zero-point values taken from the ACS Zeropoints Calculator (see Ryon
2019). Throughout this paper we will refer to the magnitudes in the
F475W, F555W, and F814W bands as, respectively, $B$, $V$, and $I$. 

Even with the shortest amongst the exposures times listed in
Table\,\ref{tab1}, some non linearity and saturation of the detector's
response is unavoidable for the brightest stars. In the NGC\,1854 field,
stars brighter than $V=16.2$ and $I=15.8$ are saturated. For NGC\,1858
saturation occurs for stars with $V \le 15.2$ and $I \le 15.0$, while in
the fields around NGC\,1856 this happens for stars with $B \le 19.3$ and
$I \le 15.7$. Nevertheless, the intrinsic brightness of these stars was
fully recovered by summing over pixels into which bleeding occurred as a
result of the over-saturation (see Gilliland 2004 and Anderson et al.
2008 for details).   

Photometric uncertainties on the magnitudes and colours of non-saturated
stars are very small (see Table\,\ref{tab2}). The magnitudes of stars
within about 2\,mag of the saturation limit are recovered with an
accuracy of typically $0.02-0.03$\,mag.

In our analysis we considered only stars with small root-mean-square
scatter in position measurements and well fitted by the ePFS routine
(Anderson et al. 2008). This sample of stars with ``high quality''
photometry was selected as in Milone et al. (2009, see their Figure\,1)
on the basis of the various diagnostics of the astrometric and
photometric quality provided by the computer programmes by Anderson et
al. (2008).

\begin{deluxetable}{llccr} 
\tablecolumns{5}
\tabletypesize{\footnotesize}
\tablecaption{Observations used in the paper.
\label{tab1}}
\tablewidth{9cm}
\tablehead{\colhead{Cluster} & \colhead{Filter} &
\colhead{Exposure time} & \colhead{Date} & \colhead{Proposal} } 
\startdata
NGC\,1854 & F555W & 50    & 2003 Oct 07 & 9891 \\
          & F814W & 40    & 2003 Oct 07 & 9891 \\
NGC\,1858 & F555W & 20    & 2003 Oct 08 & 9891 \\
          & F814W & 20    & 2003 Oct 08 & 9891 \\
NGC\,1856 F1 & F475W & $2\times665$ & 2014 Feb 09 & 13379 \\
        & F814W & $42+559$    & 2014 Feb 09 & 13379 \\
NGC\,1856 F2 & F475W & 2$\times$665 & 2014 Mar 24 & 13379 \\
        & F814W & $42+559$    & 2014 Mar 24 & 13379 \\
NGC\,1856 F3 & F475W & $2\times665$ & 2014 May 18 & 13379 \\
        & F814W & $42+559$    & 2014 May 18 & 13379 \\
NGC\,1856 F4 & F475W & $2\times665$ & 2014 Jun 06 & 13379 \\
        & F814W & $42+559$    & 2014 Jun 06 & 13379 
\enddata
\tablecomments{Exposure times are in seconds.}
\end{deluxetable}

\begin{deluxetable}{lcccccc}
\tablecolumns{7}
\tabletypesize{\footnotesize}
\tablecaption{Photometric uncertainties.
\label{tab2}}
\tablehead{\colhead{} & \multicolumn{2}{c}{NGC\,1854} &
\multicolumn{2}{c}{NGC\,1856} & \multicolumn{2}{c}{NGC\,1858} \\
\colhead{Magnitude} & 
\colhead{\hspace{0.1cm}$\sigma_{V}$} & \colhead{$\sigma_{V-I}$} &
\colhead{\hspace{0.8cm}$\sigma_{B}$} & \colhead{$\sigma_{B-I}$} &
\colhead{\hspace{0.8cm}$\sigma_{V}$} & \colhead{$\sigma_{V-I}$}}
\startdata 
     15.00 & 0.016 & 0.022 &  ---  &  ---  & 0.014 & 0.019 \\
     15.50 & 0.015 & 0.020 &  ---  &  ---  & 0.013 & 0.017 \\
     16.00 & 0.014 & 0.019 &  ---  &  ---  & 0.013 & 0.018 \\
     16.50 & 0.013 & 0.018 &  ---  &  ---  & 0.012 & 0.019 \\
     17.00 & 0.013 & 0.017 &  ---  &  ---  & 0.015 & 0.021 \\
     17.50 & 0.012 & 0.019 &  ---  &  ---  & 0.015 & 0.022 \\
     18.00 & 0.015 & 0.021 &  ---  &  ---  & 0.017 & 0.025 \\
     18.50 & 0.015 & 0.022 &  ---  &  ---  & 0.022 & 0.032 \\
     19.00 & 0.017 & 0.027 &  ---  &  ---  & 0.025 & 0.036 \\
     19.50 & 0.022 & 0.033 & 0.013 & 0.018 & 0.030 & 0.044 \\
     20.00 & 0.025 & 0.038 & 0.012 & 0.017 & 0.037 & 0.053 \\
     20.50 & 0.030 & 0.046 & 0.012 & 0.019 & 0.043 & 0.062 \\
     21.00 & 0.037 & 0.054 & 0.014 & 0.021 & 0.056 & 0.078 \\
     21.50 & 0.044 & 0.062 & 0.015 & 0.022 & 0.073 & 0.097 \\
     22.00 & 0.057 & 0.076 & 0.016 & 0.023 & 0.086 & 0.115 \\
     22.50 & 0.073 & 0.097 & 0.020 & 0.028 & 0.097 & 0.130 \\
     23.00 &  ---  &  ---  & 0.024 & 0.033 &  ---  &  ---  \\
     23.50 &  ---  &  ---  & 0.028 & 0.038 &  ---  &  ---  \\
     24.00 &  ---  &  ---  & 0.035 & 0.046 &  ---  &  ---  \\
     24.50 &  ---  &  ---  & 0.041 & 0.055 &  ---  &  ---  \\
     25.00 &  ---  &  ---  & 0.052 & 0.067 &  ---  &  ---  
\enddata
\tablecomments{Values for NGC\,1856 are representative of all fields F1--F4.}
\end{deluxetable}

\begin{figure*}
\centering
\resizebox{\hsize}{!}{\includegraphics[angle=180,origin=b,trim=0.5cm 6cm
1.5cm 2cm,clip]{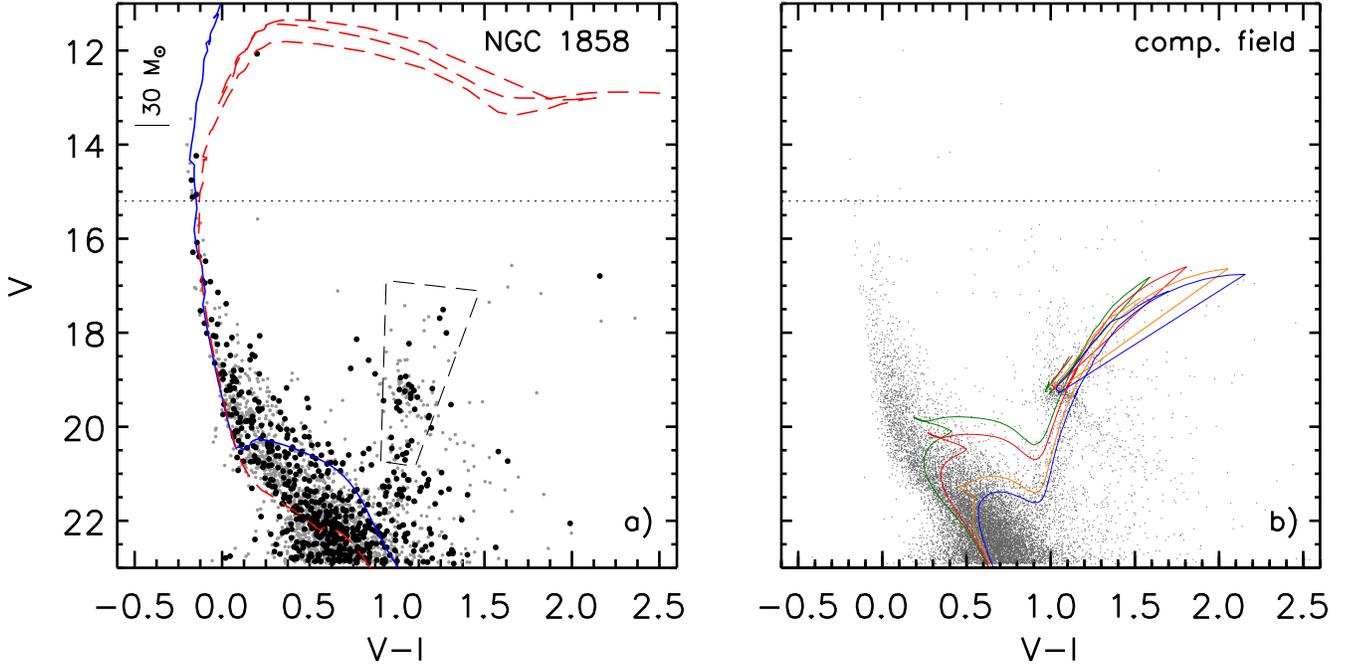}}
\caption{CMDs of high-quality stars in and around NGC\,1858. In Panel
a), small grey dots are used for stars within a radius of $12.5$\,pc of
the nominal centre (see Figure\,\ref{fig1}), while thick black dots mark
objects within $6.25$\,pc of it. The solid and dashed lines show
isochrones from the models of Chen et al. (2015) for ages of 5 and
18\,Myr, respectively, for the appropriate LMC distance, metallicity
$Z=0.007$, and a combined colour excess (foreground + intrinsic) of
$E(V-I)=0.19$. We also indicate the approximate mass of the heaviest
stars consistent with the  youngest isochrone. The CMD in Panel b) is
obtained from all stars more distant than $12.5$\,pc from the centre of
NGC\,1858 and the theoretical isochrones are from the models of Tang et
al. (2014) for metallicity $Z=0.004$ and only Galactic foreground
extinction of $E(V-I)=0.11$. Ages of $1.5$, 2, 3, and 4\,Gyr are shown,
respectively, in green, red, orange, and blue. In all panels the thin
horizontal lines indicate the saturation level discussed in
Section\,2.}  
\label{fig2}
\end{figure*}

\begin{figure*}
\centering
\resizebox{\hsize}{!}{\includegraphics[angle=180,origin=b,trim=0.5cm 6.8cm
1.5cm 2cm,clip]{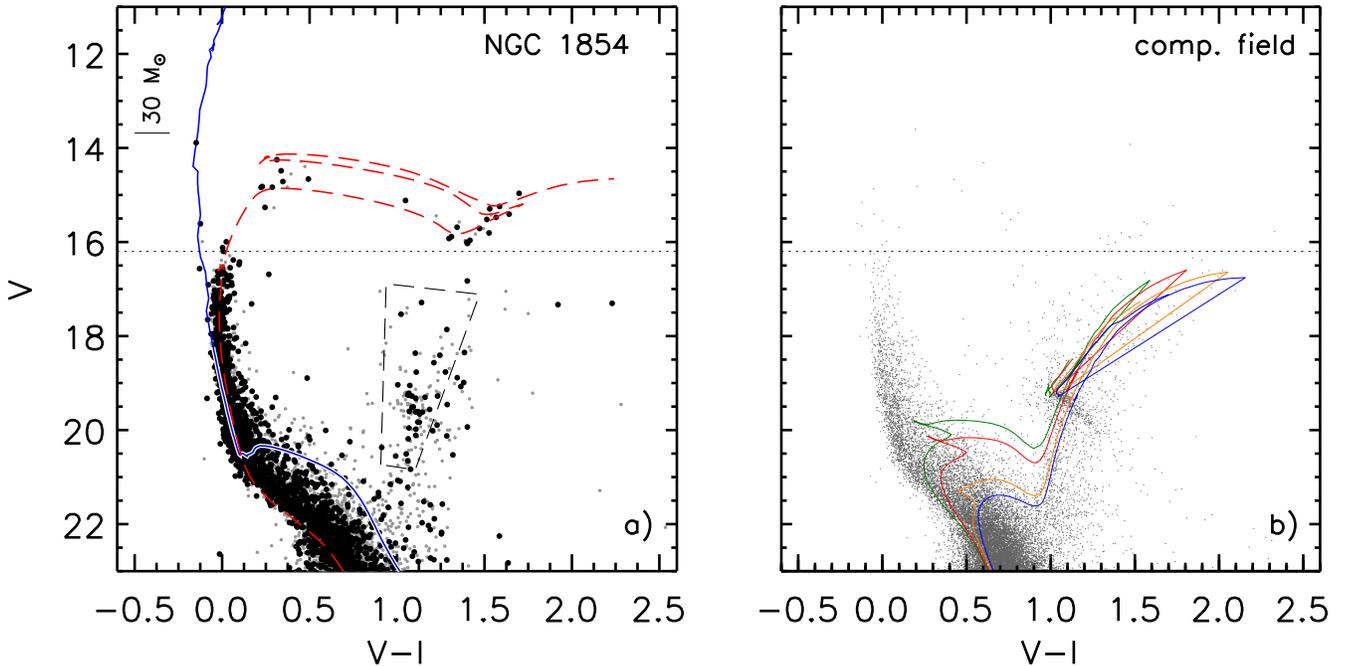}}
\caption{Same as Figure\,\ref{fig2} but for { stars in and around}
NGC\,1854. The only difference is in Panel a), where the combined colour
excess (foreground + intrinsic) applied to the isochrones is
$E(V-I)=0.21$ and the ages of the isochrones are 5 (solid line) and
60\,Myr (dashed line).}
\label{fig3}
\end{figure*}

\begin{figure*}
\centering
\resizebox{\hsize}{!}{\includegraphics{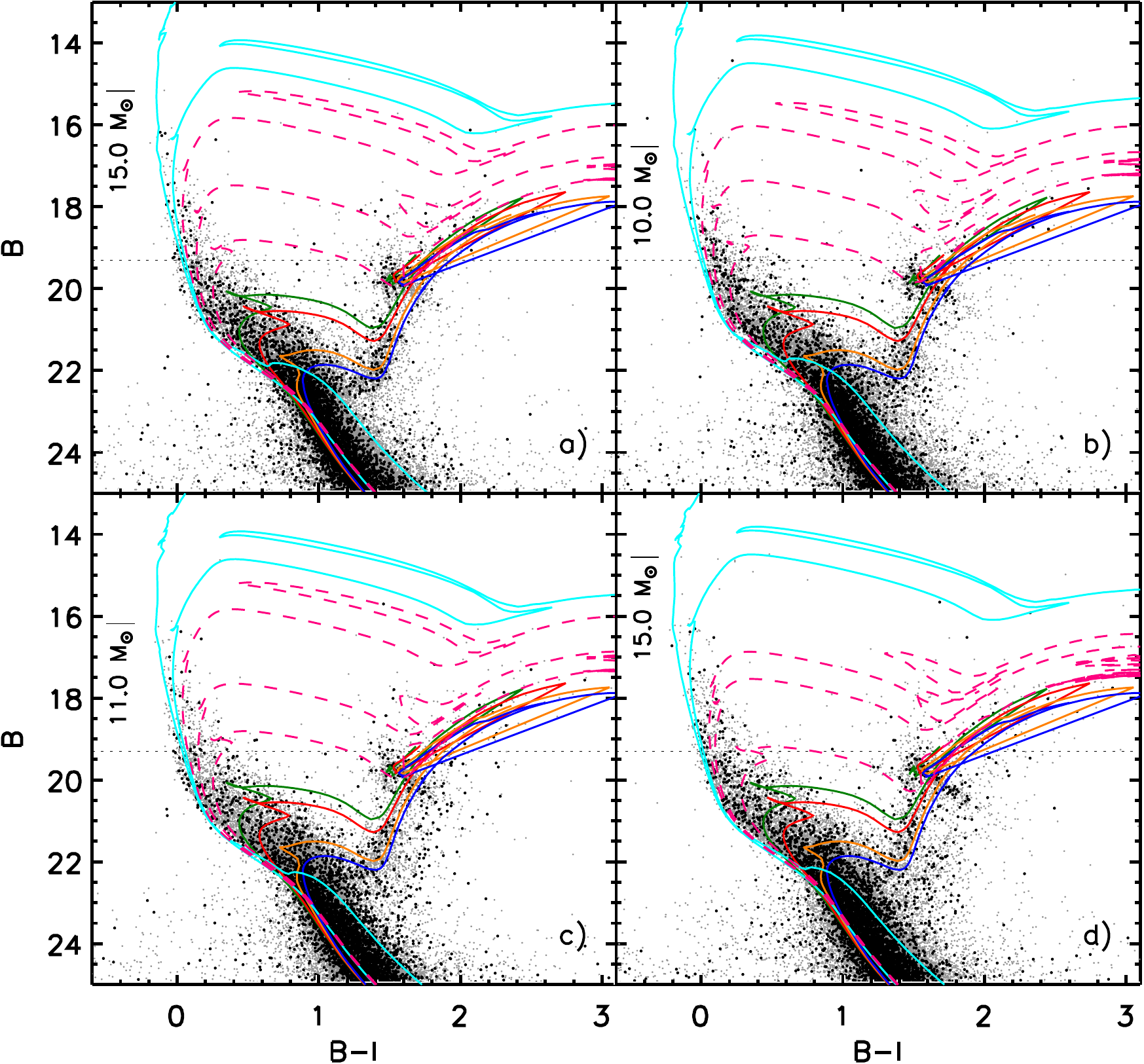}}
\caption{CMDs of the four regions around NGC\,1856 indicated in 
Figure\,\ref{fig1}. Small grey dots mark all stars, while thick black 
dots indicate objects within the circles shown in Figure\,\ref{fig1}
(radius $12.5$\,pc, for ease of comparison with Figures\,\ref{fig2} and
\ref{fig3}). Isochrones from Tang et al. (2014) for metallicity
$Z=0.004$ and ages of $1.5$, 2, 3, and 4\,Gyr (shown respectively, by
green, red, orange, and  blue thin solid lines) are the same in all
panels and only include foreground Galactic extinction $A_V=0.22$,
corresponding $E(B-I)=0.18$ for the Galactic extinction law (same as in
Figures\,\ref{fig2}b and \ref{fig3}b). Isochrones for younger ages are
from the models of Chen et al. (2015) for metallicity $Z=0.007$ and ages
and colour excess as follows. In Panel a), from Region 1, the thick
solid (cyan) {lines correspond to ages of 10 and 50\,Myr} and a 
combined colour excess (foreground + intrinsic) $E(B-I)=0.33$, while 
the dashed lines represent ages of 100, 270, and 600\,Myr with the same
combined colour excess. In Panel b), corresponding to Region 2, the ages
are of 10 and 50\,Myr for  the thick solid lines (cyan), while the
dashed lines represent ages of 120, 270, and 600\,Myr, with a combined
colour excess $E(B-I)=0.28$. In Panel c), from  Region 3, the thick
solid (cyan) lines are for ages of 12 and 60\,Myr, while the thick
dashed lines are for ages of 100, 300, and 600\,Myr, all of them with
combined colour excess $E(B-I)=0.32$. In panel d), from Region 4, the
thick solid (cyan) lines are for ages of 12 and 50\,Myr, and the thick
dashed lines for ages of 200, 300, and 800\,Myr, again with combined
colour excess $E(B-I)=0.28$.}
\label{fig4}
\end{figure*}

\vspace*{0.5cm}
\section{Colour--magnitude diagrams: multiple populations}

We show in Figures\,\ref{fig2} and \ref{fig3} the CMDs obtained from the
stars with high-quality photometry in the fields { including and
surrounding} NGC\,1858 and NGC\,1854, respectively. The CMDs reveal a complex
population, made up of stars in different evolutionary phases. In Panel
a) of both figures small grey dots sample a region of $12.5$\,pc radius
around the nominal centres of the two clusters (dashed circles in
Figure\,\ref{fig1}). Objects within the inner $6.5$\,pc are indicated
with thick black dots. Both regions reveal a young population of stars
in the upper main sequence (MS) and a sparsely populated red giant
branch (RGB). Also shown are theoretical isochrones from the models of
Chen et al. (2015) for metallicity $Z=0.007$, since this is a typical
value for the LMC (e.g., Hill et al. 1995; Geha et al. 1998). All
isochrones already include a distance modulus $(m-M)_0=18.55$ (Panagia
1998) for the LMC and the reddening contribution of the Milky Way along
the line of sight to these clusters, namely $A_V=0.22$ (e.g.,
Fitzpatrick \& Savage 1984), which corresponds to $E(B-V)=0.07$ and
$E(V-I)=0.11$ for the extinction law of the diffuse Galactic ISM (e.g.,
Cardelli et al. 1989; Fitzpatrick \& Massa 1990). 

The best fit to the upper MS of NGC\,1858 is obtained for an age of 5
Myr (blue solid line in Figure\,\ref{fig2}a) and requires an extra
$E(V-I)=0.08$ component of color excess in addition to the foreground
$E(V-I)=0.11$ mentioned above,\footnote{As we will show in Section 5, in
all these regions also the ratio $R_V$ between total and selective
extinction is considerably higher than the characteristic $R_V=3.1$
value typical of the diffuse Galactic ISM.} thus in total $E(V-I)=0.19$.
With the same total reddening, the blue supergiant HD\,261196 at
$V=12.07$, $V-I= 0.20$ is compatible with an age of $\sim 13$\,Myr (red
dashed line). We note that for this object we have used the magnitudes
published by Bonanos et al. (2009) because in our ACS images the star is
saturated. 

Concerning NGC\,1854, the best fit to its upper MS is also obtained for
an age of 5 Myr (blue solid line in Figure\,\ref{fig3}a) and a slightly
larger value of the colour excess, namely $E(V-I)=0.21$. Adopting the
same extinction value also for the supergiants located near $V=14$,
$V-I=0.3$ suggests an age around 60\,Myr for these objects (red dashed
line). 

The CMDs of both clusters (Figures\,\ref{fig2}a and \ref{fig3}a) also
reveal a sparsely populated RGB, which however is consistent with
contamination by { LMC} field stars { along the line of sight
falling} within the selected radius. To explore
this, we counted the number of stars inside the dashed areas of the CMDs
in Figures\,\ref{fig2}a and \ref{fig3}a and compared them with the
number of objects in identical regions of the CMDs of the comparison
fields. The latter are shown in Figures\,\ref{fig2}b and \ref{fig3}b,
and include all stars farther than $12.5$\,pc from the centres of
NGC\,1858 and NGC\,1854, respectively. Once scaled by the relative areas
spanned by each cluster and its comparison field, the number of stars
within the dashed trapezoids in the CMDs of the cluster and of the field
are indisinguishable, within statistical uncertainties. 

Furthermore, objects in the RGB phase would not be compatible with the
young ages of both clusters, in the range 5 -- 60\,Myr derived above,
requiring instead ages in excess of $\sim 1$\,Gyr. This is shown
graphically in Figures\,\ref{fig2}b and \ref{fig3}b, where { over}
the CMDs of the { comparison} fields { surrounding} NGC\,1858 and
NGC\,1854 { we show} theoretical isochrones for ages of $1.5$, 2, 3,
and 4\,Gyr, respectively in green, red, orange, and blue. The isochrones
are taken from the models of Tang et al. (2014) for metallicity
$Z=0.004$ and already include Galactic foreground extinction of
$E(V-I)=0.11$. { We do not include any additional colour excess with
these isochrones since old LMC field stars can be located anywhere along
the line of sight, not only behind but also in front of the young
clusters, whose exact position is not known. We also note that, in
principle, red clump stars in this field could have ages up to  10 Gyr.
However, as Girardi \& Salaris (2000) pointed out, the age  distribution
of red clump stars in galaxies with constant star formation is strongly
skewed towards younger ages, due to the longer lifetimes of  more
massive RC stars and to the decreasing rate at which stars leave the  MS
at older ages. Moreover, as pointed out in De Marchi et al. (2014),  RC
stars with ages between 3 and 9 Gyr change very little their intrinsic 
positions in the CMDs. This justifies limiting our interval of ages up
to  $\sim 4$ Gyr. } 

Besides a broad MS, the prominent RGB is the characteristic feature of
these CMDs, together with a remarkably elongated RC, extending by over
one magnitude in $V$ and suggesting a considerable amount of
differential reddening in these fields. The extended RC is clearly not
caused by age effects, since all isochrones agree with the overdensity
oberved in the CMD at $V-I \simeq 1.0$, $V\simeq19.2$ (the nominal RC
location), but not with the extended tail. { More details on the
effects of a range of ages on the shape and extent of the RC in the CMD
are provided by De Marchi et al. (2014). In particular, for the
metallicity $Z=0.004$ that is relevant to this work, their Figure\,4
shows that the combination of stellar populations with ages ranging from
$1.4$ to 3\,Gyr only causes a broadening of less than $0.05$ and
$0.02$\,mag (1\,$\sigma$) respectively on the $V$ magnitude and $B-V$
colour of the resulting RC ($\sigma=0.02$\,mag is also the value found 
in the $V-I$ colour). Considering that the nominal RC position
corresponding to each individual age already has an intrinsic spread of
$\sim 0.1$\,mag in $V$ and $\sim 0.05$\,mag in $B-V$, the broadening
introduced by the age spread is marginal.}

The CMDs of the third region studied in this work, around NGC\,1856, are
shown in Figure\,\ref{fig4}. As mentioned in Section\,2, we selected
observations of four fields { surrounding (but not containing)} the 
cluster, at a typical distance of $5^\prime$ or about 75\,pc from the
cluster centre (see Figure\,\ref{fig1}). In these regions no
concentrations of objects or clustering of bright stars are seen,
indicating that these areas are dominated by { LMC} field stars. In
Figure\,\ref{fig4} we show the CMDs of the four fields, using small
grey  dots to indicate all objects and thick black dots for stars within
the circular regions shown in Figure\,\ref{fig1}. The circular regions,
with a radius of $12.5$\,pc, have been selected to simplify the
comparison with the CMDs of { the fields around} NGC\,1858 and
NGC\,1854, but they do not correspond to any overdensity of objects, as
mentioned above.

Most stars in these fields are old, as indicated by the prominent RGB,
for which comparison with isochrones suggests ages older than $\sim
1$\,Gyr. The green, red, orange, and blue isochrones are the same as
shown in Figures\,\ref{fig2}b and \ref{fig3}b from Tang et al. (2014)
for ages of $1.5$, 2, 3 and 4\,Gyr, metallicity $Z=0.004$, and already 
include Galactic foreground extinction, which in these bands amounts to 
$E(B-I)=0.18$. 

Besides old stars, a much younger population is also present in these
CMDs, as witnessed by the many upper MS stars. These objects are
compatible with ages in the range $\sim 10 - 50$\,Myr (for details see
caption to Figure\,\ref{fig4}) and masses up to $\sim 10-15$\,\Msolar.
Furthermore, a number of objects in the range $0.6 \la B-I \la 1.6$ and 
$16 \la B \la 19$ are consistent with giants with ages in the range
$\sim 100 - 600$\,Myr. Thus, in spite of the lack of clear clustering
or overdensities, in these regions star formation has been
proceeding in the past several 10\,Myr and in the past several
100\,Myr. This finding will be crucial for understanding the extinction
properties in these fields (see Section\,5).

Like in the case of { the fields around} NGC\,1858 and NGC\,1854,
also here does the RC show an extended shape, suggesting the presence of
patchy extinction. The extent of the RC elongation is not the same in
all fields and increases proceeding from Panel a) to d). In the
following section we will study the extinction properties in all these
fields through an analysis of the shape of the extended RC.


\begin{figure*}
\centering
\resizebox{\hsize}{!}{\includegraphics{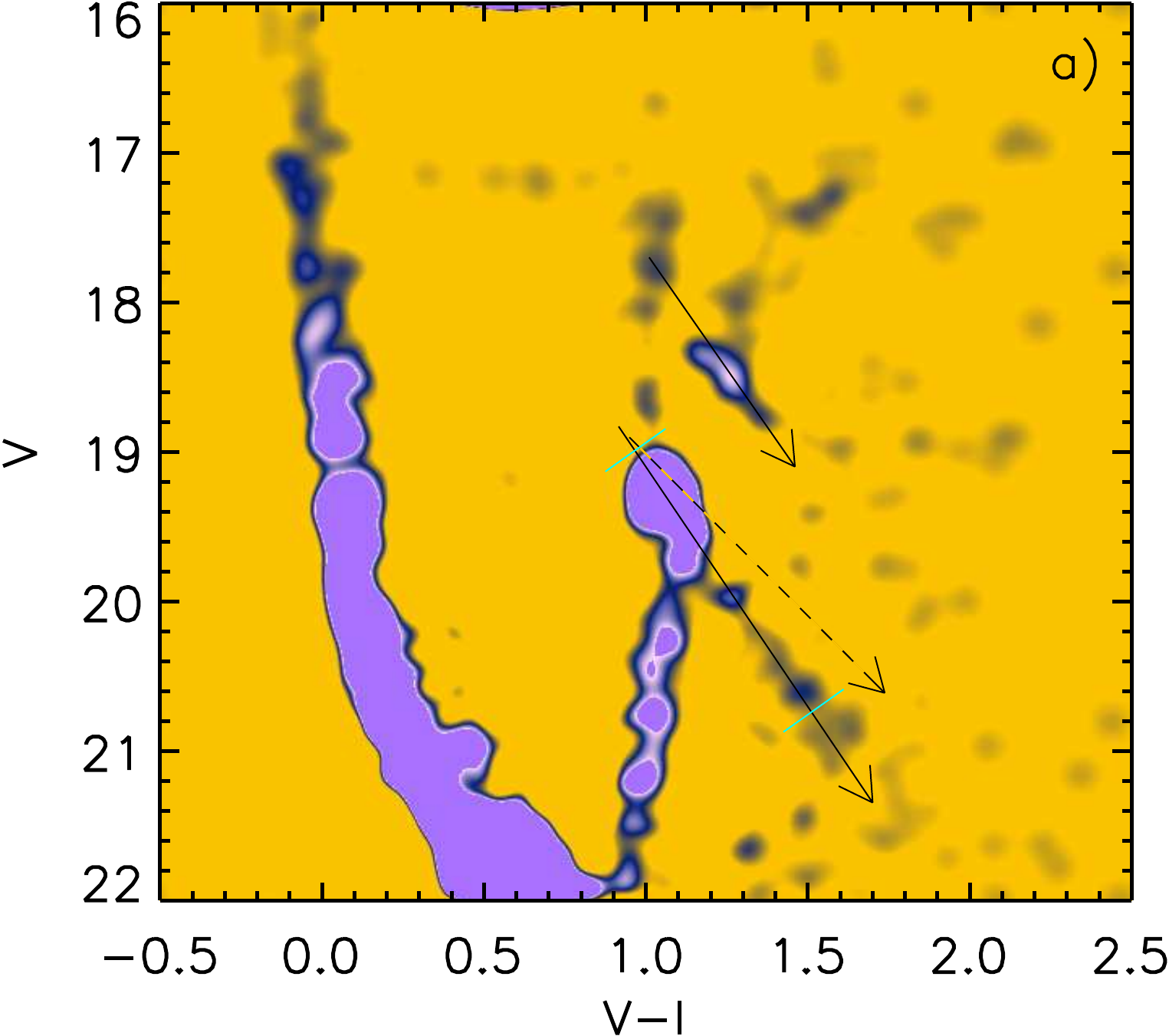}
                      \includegraphics{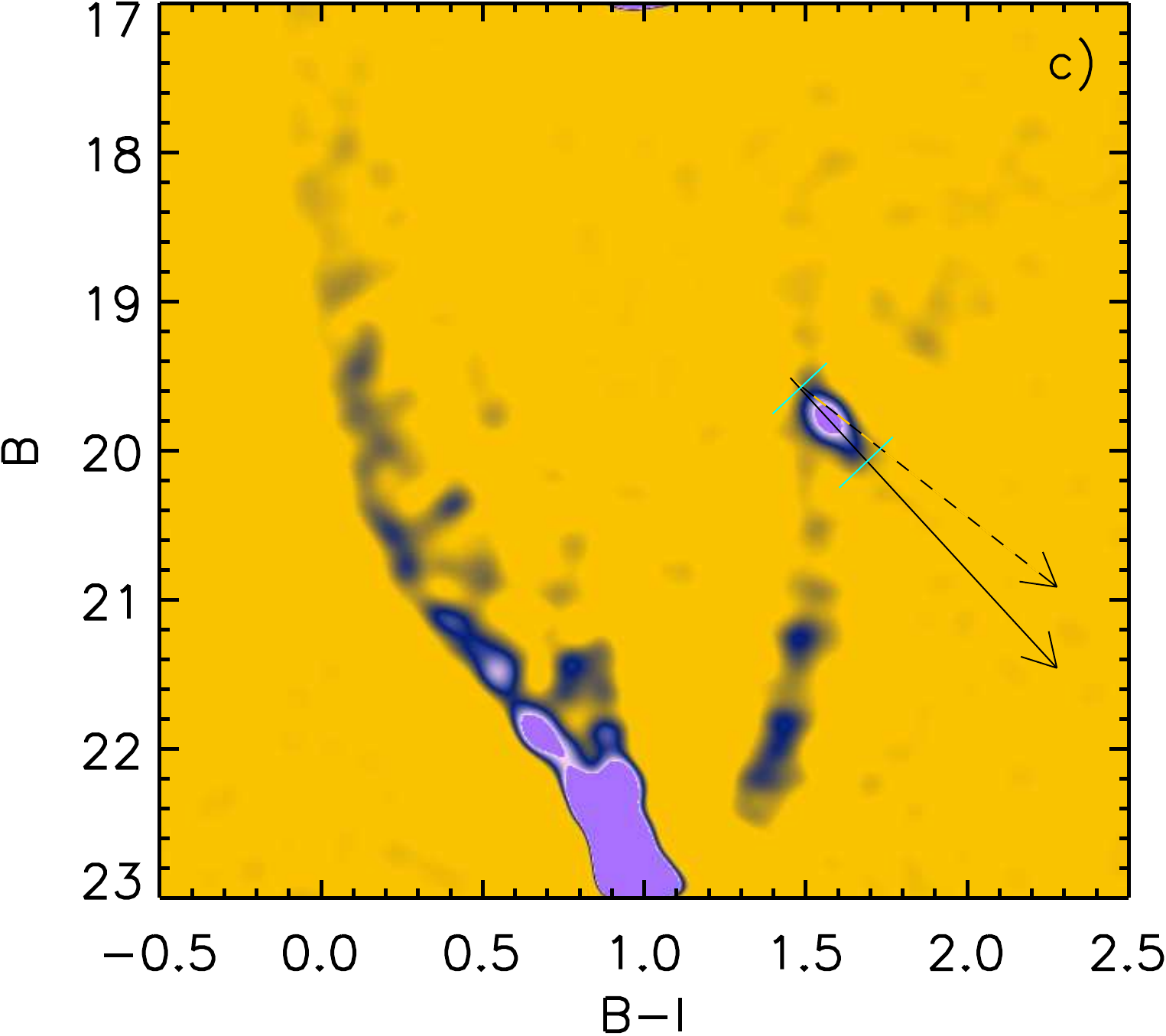}
                      \includegraphics{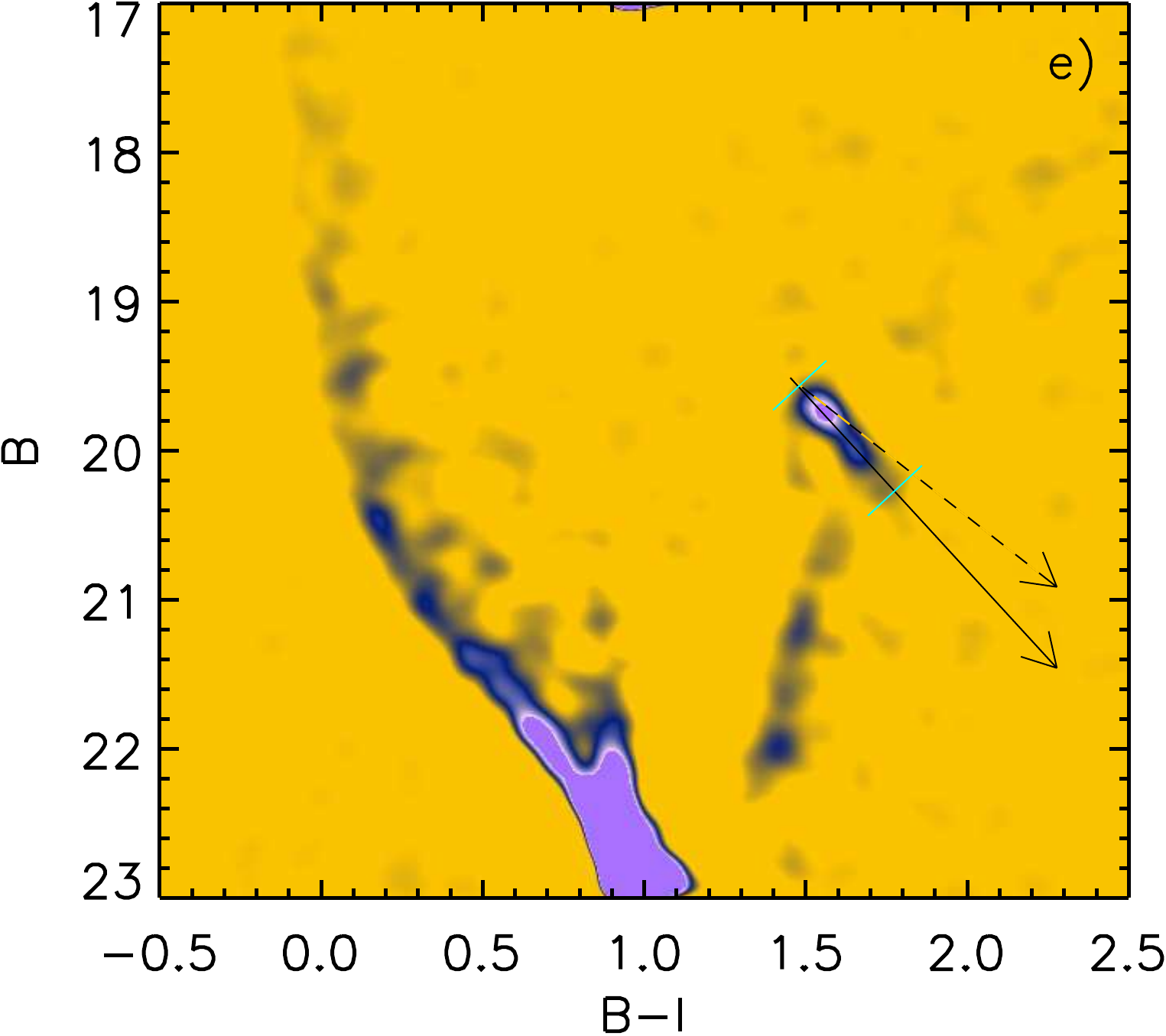}}
\resizebox{\hsize}{!}{\includegraphics{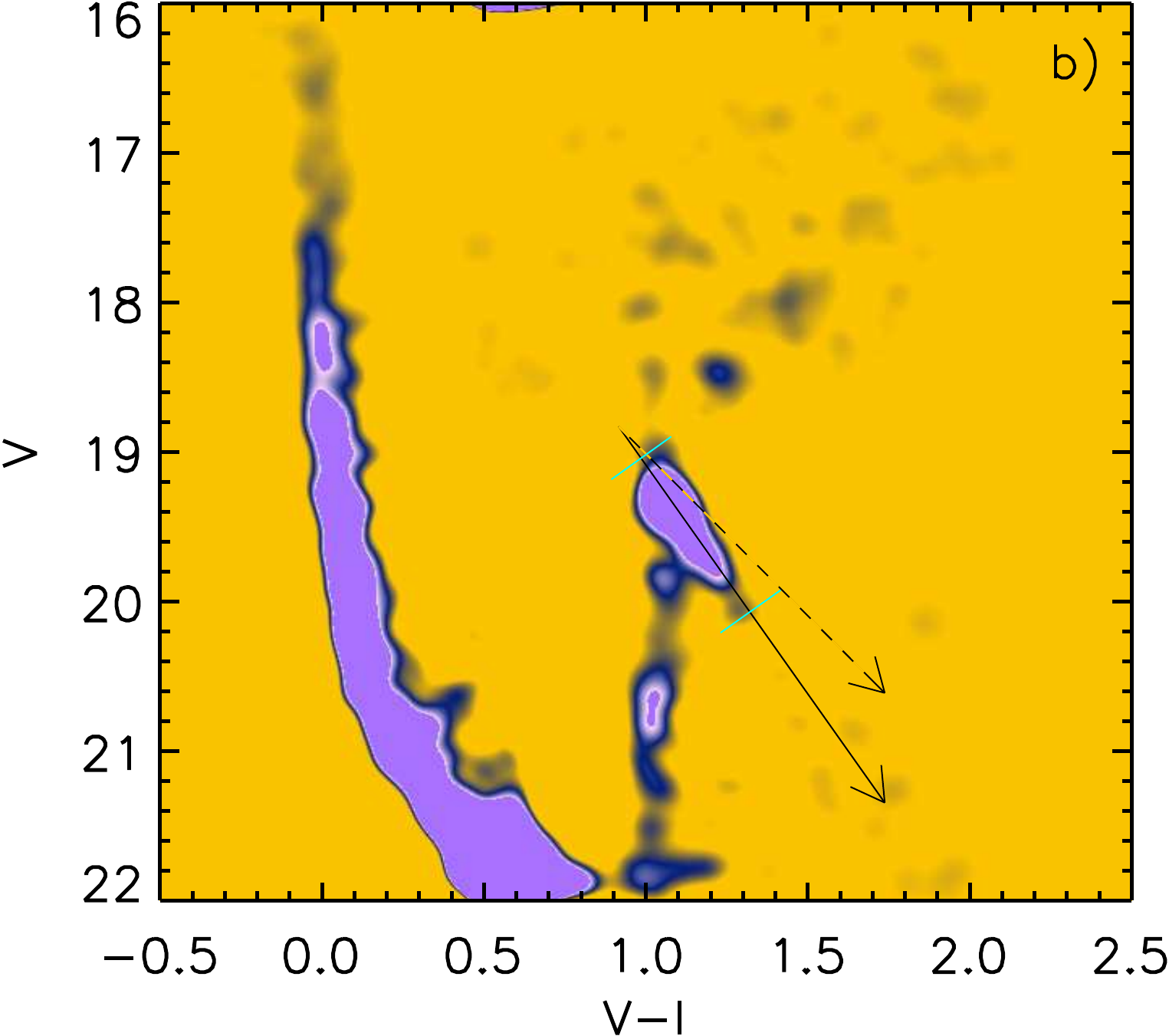}
                      \includegraphics{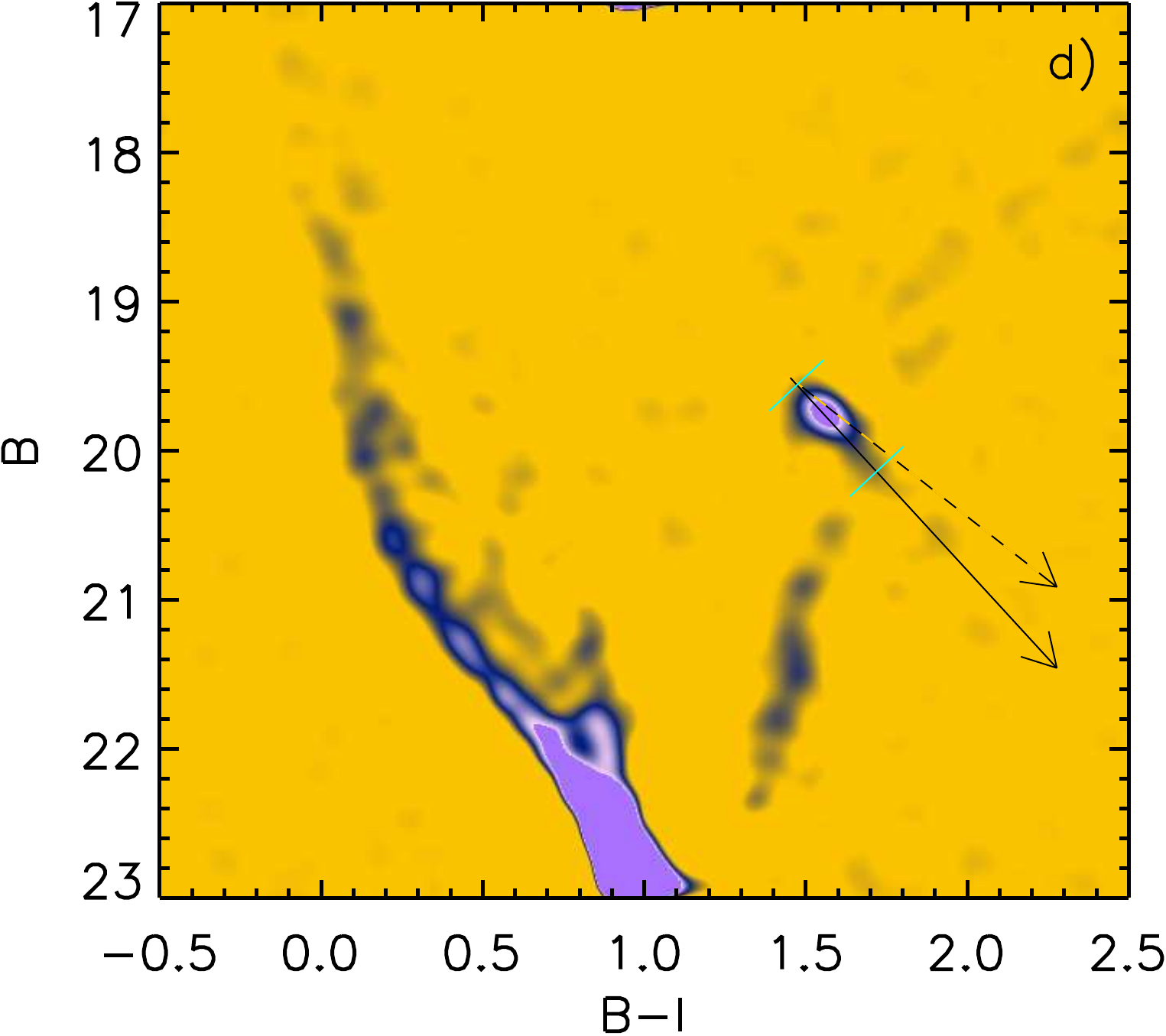}
                      \includegraphics{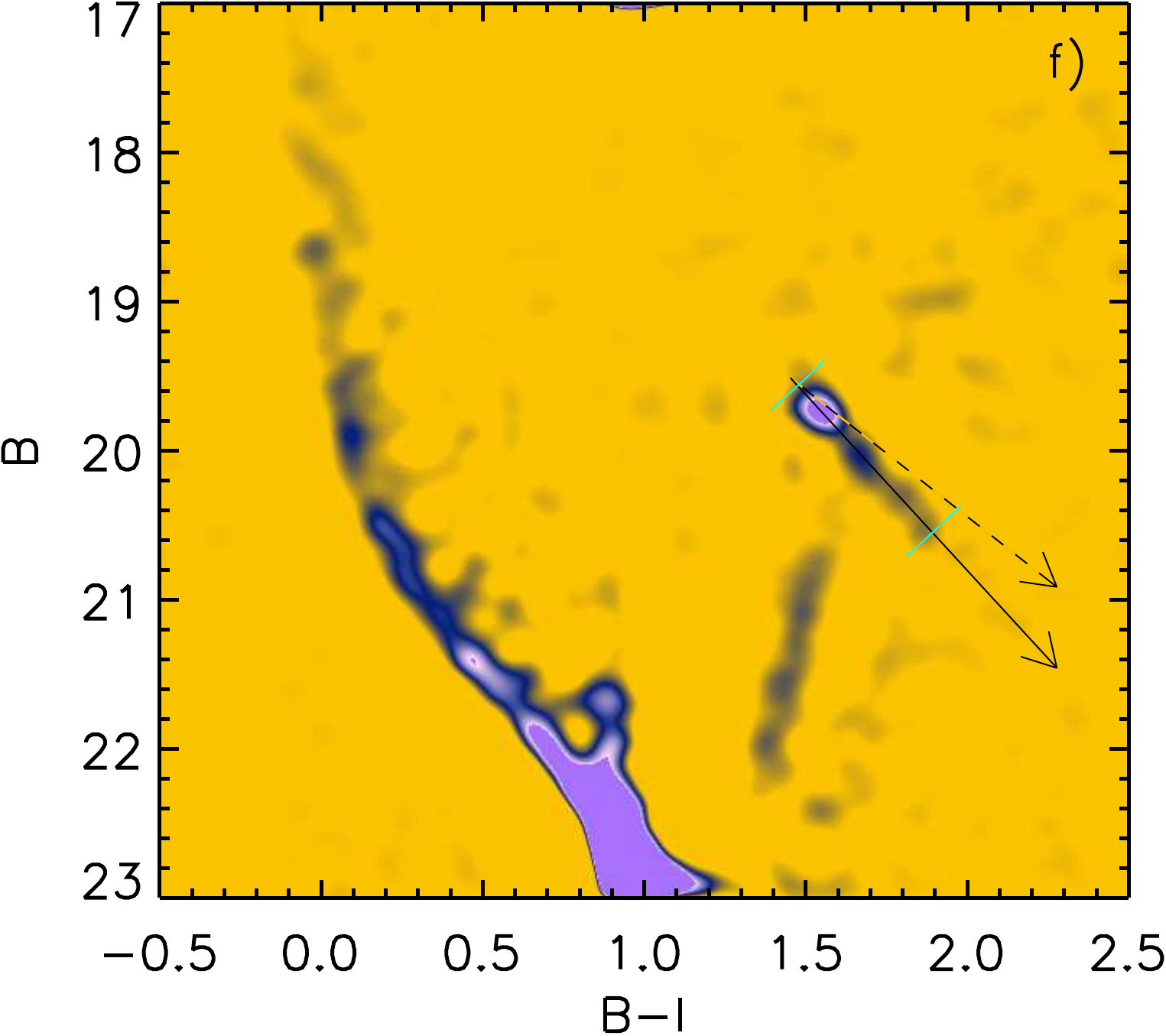}}
\caption{Unsharp masking applied to all CMDs. { Panels are as
follows:  a) NGC\,1858; b) NGC\,1854; c) NGC\,1856 F1; d) NGC\,1856 F2;
e)  NGC\,1856 F3; f) NGC\,1856 F4.} The slope of the reddening vector
is  obtained through a linear fit to the elongated RC. The extent of
the  elongated RC is indicated by the tick marks. { The short solid
arrow starting at  $V\simeq 17.6$ in panel a) is parallel to the
reddening vector for this field and provides a good fit to the younger 
extended RC.} }
\label{fig5}
\end{figure*}

\vspace*{1cm}
\section{Red clump stars to probe extinction}

The stability associated with the phase of central He-burning
characteristic  of RC stars (e.g. Cannon 1970) makes these objects
excellent tracers of reddening, when the distance is known. Girardi \&
Salaris (2001) and Salaris \& Girardi (2002) presented a detailed study
of the properties of the mean RC as a function of age, metallicity, and
star-formation history. As shown in Figures\,\ref{fig2}, \ref{fig3}, and
\ref{fig4}, the expected location of the RC in the CMD is only
marginally affected by age differences as large as $2.5$\,Gyr. To be
sure, when objects with different ages and metallicities are present in
the stellar population, the RC does take on a slightly elongated shape,
but De Marchi et al. (2014) showed that for LMC stars these effects
account for a dispersion of at most $0.2$\,mag in the $V$ and $I$ bands
when both the age and metallicity vary by a factor of two. 

Also considerable dispersion in distance along the line of sight could
of course cause the RC to appear elongated in the CMD, { albeit only
vertically, along the magnitude axis, since distance does not have any
effect on the colour of stars. Moreover,} for stars in the LMC this
effect is { negligible}, since that galaxy is seen at a high
inclination ($\sim 35^\circ$) and its disc has a scale height of
typicaly less than $0.5$\,kpc (Van der Marel \& Cioni 2001), which is
not significant when compared with the distance to the LMC itself ($51.4
\pm 1.2$\,kpc; Panagia et al. 1991, and updates in Panagia 1998, 1999).
{ Indeed, an exponential distribution with a scale height of
$0.5$\,kpc at the distance of the LMC results in a median deviation
along the line of sight of less than $0.02$\,mag even including the
thickening caused by the $35^\circ$ inclination. For 83\,\% of the stars
the deviation is less than $0.05$\,mag. The effect of distance along the
line of sight is, therefore, negligible. Even when combined with the
small broadening of the RC caused by age differences, these effects
remain barely detectable and cannot account for the substantial RC 
elongation in colour and magnitude that we observe in these regions.}

In this work we are interested not only in the elongation of the RC,
which is a measure of the total extinction in the field, but also in the
slope of the extended RC in the CMD, since this gives a fully empirical
measure of the direction of the reddening vector (Nataf et al. 2013; De
Marchi et al. 2014). Together, the two quantities provide information 
on the extinction properties in the field. 

A practical method to measure both the length and the slope of the RC
feature in the CMD is to apply the unsharp-masking technique. De Marchi
et al. (2016) provide a detailed description of the method, which we
briefly summarise here. The purpose of unsharp masking is to make an
image of the CMD sharper by subtracting from it a mask consisting of a
blurred version of the CMD image itself. First, to obtain the image of
the CMD, each object in it is mapped to a two-dimensional array, with a
sampling of $0.01$ mag in colour and magnitude. The array is then
convolved with a narrow Gaussian beam, which assigns to each point in
the CMD the resolution pertaining to the photometric uncertainties. We
used a beam size $\sigma=0.08$\,mag, corresponding to about three times
the typical photometric uncertainty. Similarly, to obtain the blurred
mask, the same array is convolved with a wider Gaussian beam, in our
case $\sigma=0.3$\,mag. The mask is then subtracted from the CMD image.
Analytically, these operations are equivalent to convolving the CMD with
a kernel represented by the difference between two Gaussian beams with
different $\sigma$ (see De Marchi et al. 2016). { We experimented
with different values for the wider Gaussian beam (namely $0.2$ and
$0.4$\,mag) and the differences are imperceptible.}

We show in Figure\,\ref{fig5} the CMDs of the various regions after
unsharp masking. { Panels a) and b) refer to the regions surrounding
NGC\,1858 and NGC\,1854, respectively, while Panels c) through to f)
refer to fields F1 to F4 around NGC\,1856.} The high-frequency
substructures are easier to identify than in the original CMDs, in
particular the MS, the MS turn-off, the RG branch, and of course the
elongated RC. The direction of the reddening vector (thick solid arrows
in Figure\,\ref{fig5}) is measured using the ridge line of the extended
RC. The uncertainty on the slope is obtained using weights proportional
to the local density of points in the CMD. In Figure\,\ref{fig5}, the
dashed lines show the reddening  vector corresponding to the extinction
law of the Galactic diffuse ISM, in the bands specific to each panel. It
is evident that in these regions the slope is considerably steeper than
in the Galactic ISM, about $1.5$ times as steep.

The slope of the reddening vector corresponds in turn to the ratio
between total ($A$) and selective ($E$) extinction in these specific
bands. The values measured in the individual regions are given in
Table\,\ref{tab3}, together with their uncertainties. We also show the
values measured in the same bands in and around 30 Dor and in NGC\,1938,
together with their uncertainties, as well as the values corresponding
to the average extinction law in the diffuse Galactic ISM. The
uncertainty on the latter does not reflect the actual measurement
errors, but rather the wide dispersion around the mean, exceeding 20\%
(see e.g. Herbst 1975; Massa et al. 1983; Fitzpatrick \& Massa 2005;
Nataf et al. 2016).

{ We note in passing that also the feature seen in
Figure\,\ref{fig5}a at $1.0<V-I< 1.4$ and $18.3 <V< 18.9$ is an
elongated RC. Comparison with the Chen et al. (2015) isochrones
mentioned above suggests that this RC is associated to a population of
intermediate age, $\sim 330$\,Myr, and as such considerably younger than
the population responsible for the main RC feature at fainter
magnitudes. According to the isochrones, the unextinguished location of
the RC for this population should be at $V=17.8$ and $V-I=1.0$, where an
enhancement is indeed present in the CMD, both before and after unsharp
masking. The slope of the younger extended RC is consistent with the
reddening vector measured elsewhere in this field (see short solid
arrow). Its shape and appearence suggests that most of these RC stars
are behind the NGC\,1858 cluster, because of the gap that separates the
nominal RC position at $V=17.8$ from the blue end of the elongation at
$V=18.2$. The implied minimum reddening value of $A_V=0.4$ is  larger
than the minimum intrinsic reddening revealed by the massive stars in 
the upper MS of NGC\,1858, which is of the order of $A_V=0.25$ (see
Figure\,\ref{fig2}a). }

\begin{deluxetable}{lcc} \tablecolumns{3}
\tabletypesize{\footnotesize}
\tablecaption{Ratio of total and selective extinction in our regions.
For comparison, we also indicate the values in the same bands
in and around 30 Dor, as well as in the diffuse Galactic ISM . 
\label{tab3}}
\tablehead{\colhead{Region} & \colhead{$A_V/E(V-I)$} &\colhead{$A_B/E(B-I)$} \\[0.05cm]
\multicolumn{1}{c}{(1)} & \multicolumn{1}{c}{(2)} &
\multicolumn{1}{c}{(3)}}
\startdata
  NGC 1858 & $3.21 \pm 0.14$ & \\    
  NGC 1854 & $3.26 \pm 0.35$ & \\
  NGC 1856 field 1 &  & $2.70 \pm 0.53$ \\
  NGC 1856 field 2 &  & $2.31 \pm 0.26$ \\
  NGC 1856 field 3 &  & $2.29 \pm 0.28$ \\
  NGC 1856 field 4 &  & $2.33 \pm 0.12$ \\
  NGC 1938 & $2.99 \pm 0.12$ & \\
  30 Dor   &  $2.97 \pm 0.08$ & $2.21 \pm 0.14$ \\
  30 Dor West (NGC 2060) & $3.17 \pm 0.29$ & $2.41 \pm 0.31$ \\ 
  & & \\
  Diffuse Galactic ISM& $2.17\pm0.44$ & $1.70\pm0.34$ 
\enddata
\end{deluxetable}


\section{Discussion}

Table\,\ref{tab3} reveals that the reddening vectors for the { lines
of sight towards the three clusters and their surroundings} are
systematically steeper than { those towards} NGC 1938 or 30 Dor
(where $R_V=4.5\pm0.2$; De Marchi \& Panagia 2014). The slopes are more
similar to those measured { towards} NGC\,2060 (30 Dor West), where
the extinction law derived by De Marchi et al. (2014) corresponds to
$R_V=5.6 \pm 0.3$. This suggests that { the lines of sight towards}
NGC\,1854, NGC\,1856, and NGC\,1858 share similar extinction properties
to those { towards} 30 Dor West, with a value of $R_V\simeq5.5$. 

Similarly to 30 Dor and 30 Dor West, the likely reason for the elevated
value of $R_V$ in { the direction of} NGC\,1854, NGC\,1856, and
NGC\,1858 is the presence of a grey component, caused by an extra
population of big grains, superposed to the more standard LMC extinction
curve (Gordon et al. 2003). In 30\,Dor, De Marchi \& Panagia (2019)
showed that the effect of the  additional grey component is present at
all wavelengths shorter than  $\sim 1\,\mu$m and through to the far
ultraviolet range. That work reveals that the big grains responsible for
the grey component { towards}  30\,Dor are of the same nature as
those present in the diffuse ISM of the MW, but their fraction is about
twice as high. The higher value of $R_V$ { towards} the three
clusters studied here suggests possibly an even larger fraction of big
grains.

We highlight that the extinction properties in these fields, as probed
by the slope of the reddening vector, do not correlate with the amount
of extinction, which is indicated by the length of the extended RC. All
fields have a slope consistent with $R_V\simeq 5.5$ but the range of
extinction values across these fields varies considerably. This can be
seen in Figure\,\ref{fig5}, where the extent of the RC in each CMD is
marked by the short segments. The marks correspond to the farthest
points along  the RC where the density of stars exceeds the 95
percentile ($2\,\sigma$) and provide a measure of the spread of
extinction present in the fields. The bluest end of the RC corresponds
to no intrinsic extinction (and hence with $A_V\simeq 0.2$ once the
Galactic foreground component is taken into account), while the most
reddened end, { towards} NGC\,1858, corresponds to $A_V\simeq1.7$ or
total extinction $A_V \simeq 1.9$ once also the contribution of the
Milky Way is included. 

The range of extinction values is derived from the magnitude difference 
between the marks, taking into account the intrinsic size of the
undispersed RC, which in these bands amounts to $\sim 0.1$\,mag (Girardi
\& Salaris 2001). {We underline that if instead of adopting
$\sigma=0.3$\,mag as the size of the smoothing Gaussian beam we had used
$0.2$\,mag or $0.4$\,mag the resulting length of the RC would have been,
respectively, $0.02$\,mag shorter or $0.01$\,mag longer than the $A_V
\simeq 1.9$ value given above. These differences are negligible when
compared to the the intrinsic size of the undispersed RC.}

{ As already mentioned in Section\,4,} De Marchi et
al. (2014) showed that a spread of a factor of two on both the age and
metallicity of the stars only broadens the size of the RC to about
$0.2$\,mag in $V$ and $I$. Because of the differential way in which they
are derived, the measured extinction spreads are not affected by the
Galactic foreground extinction along the line of sight. 

Using the same definition of the extinction spread for all CMDs in
Figure\,\ref{fig5} allows us to compare to one another the ranges of
extinction in the different regions. They are shown in Table\,\ref{tab4}
for the $B$, $V$, and $I$ bands, respectively indicated as $\Delta A_B$,
$\Delta A_V$, $\Delta A_I$. The values of $\Delta A_B$ are directly
measured { towards} NGC\,1856 in Figure\,\ref{fig5}, while those of
$\Delta A_V$ are measured in the same figure { along the lines of
sight to} NGC\,1854 and NGC\,1858. The extinction law { towards} 30
Dor West (De Marchi et al. 2014) implies $A_B=1.2 \, A_V$ and this
relationship was used here to transform the measured $\Delta A_B$ values
into $\Delta A_V$, and viceversa (derived quantities are shown in
Italics in Table\,\ref{tab4}). All values of $\Delta A_I$ are measured
directly for all regions from CMDs similar to those of
Figure\,\ref{fig5} but in which the $I$ magnitude is plotted as a
function of the $B-I$ and $V-I$ colours. 

\begin{deluxetable}{lcccc} \tablecolumns{3}
\tabletypesize{\footnotesize}
\tablecaption{Total extinction in each field, in various bands, compared
with the number of stars more massive than 8\,\Msolar. 
\label{tab4}}
\tablehead{\colhead{Region} & \colhead{$\Delta A_B$}
&\colhead{$\Delta A_V$} & \colhead{$\Delta A_I$} & \colhead{$N$}
\\[0.05cm]
\multicolumn{1}{c}{(1)} & \multicolumn{1}{c}{(2)} &
\multicolumn{1}{c}{(3)} & \multicolumn{1}{c}{(4)} & 
\multicolumn{1}{c}{(5)}}
\startdata
  NGC 1856 field 1 & $0.39$ & {\em 0.31} & $0.18$ & $6$ \\
  NGC 1856 field 2 & $0.48$ & {\em 0.38} & $0.23$ & $4$ \\  
  NGC 1856 field 3 & $0.45$ & {\em 0.36} & $0.22$ & $4$ \\  
  NGC 1856 field 4 & $0.88$ & {\em 0.72} & $0.47$ & $12$ \\
  NGC 1854     & {\em 1.14} & $0.93$     & $0.59$ & $11$ \\
  NGC 1858     & {\em 1.94} & $1.60$     & $1.05$ & $20$ 
\enddata
\end{deluxetable}


\begin{figure*}
\centering
\resizebox{\hsize}{!}{\includegraphics{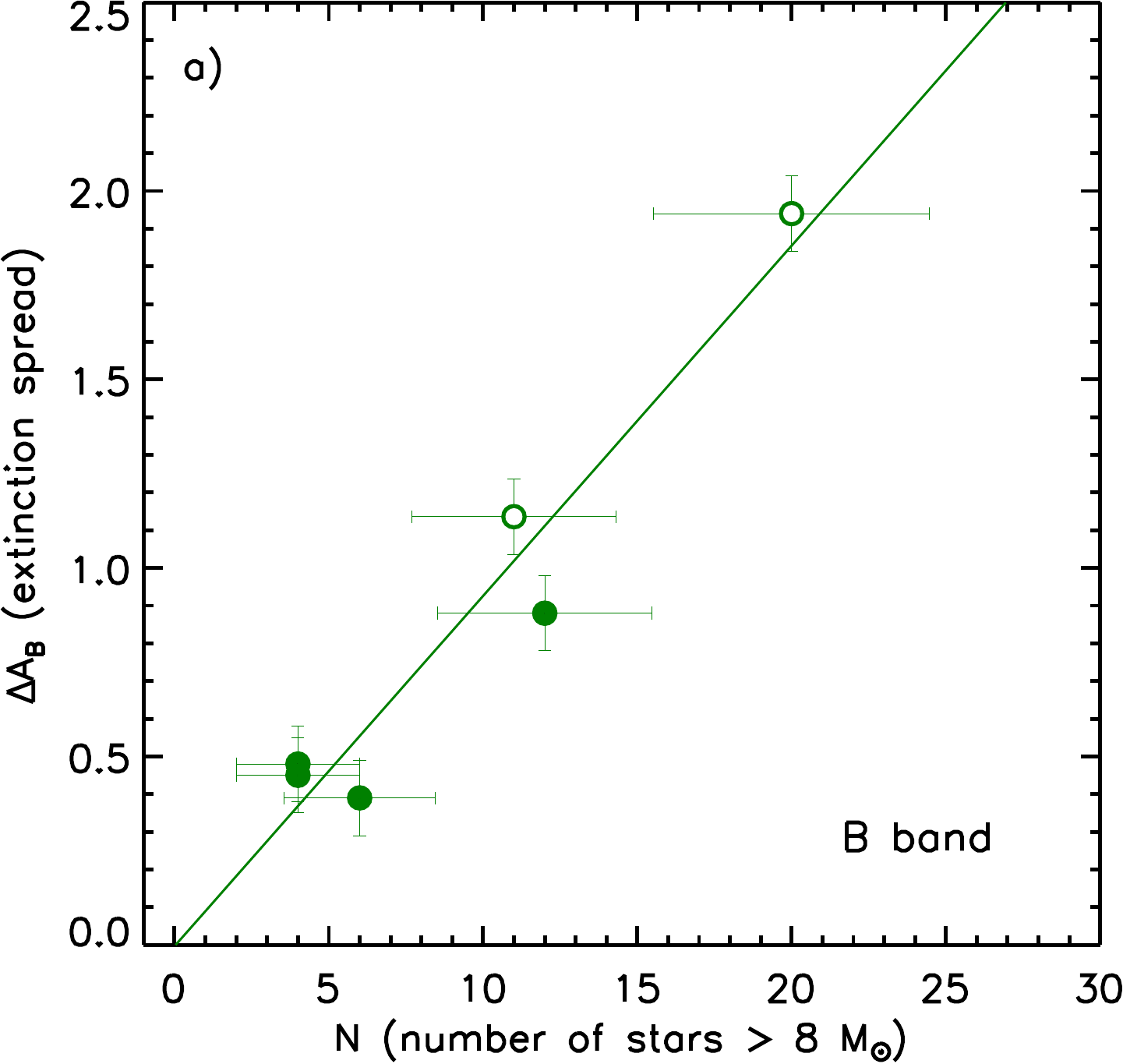}
                      \includegraphics{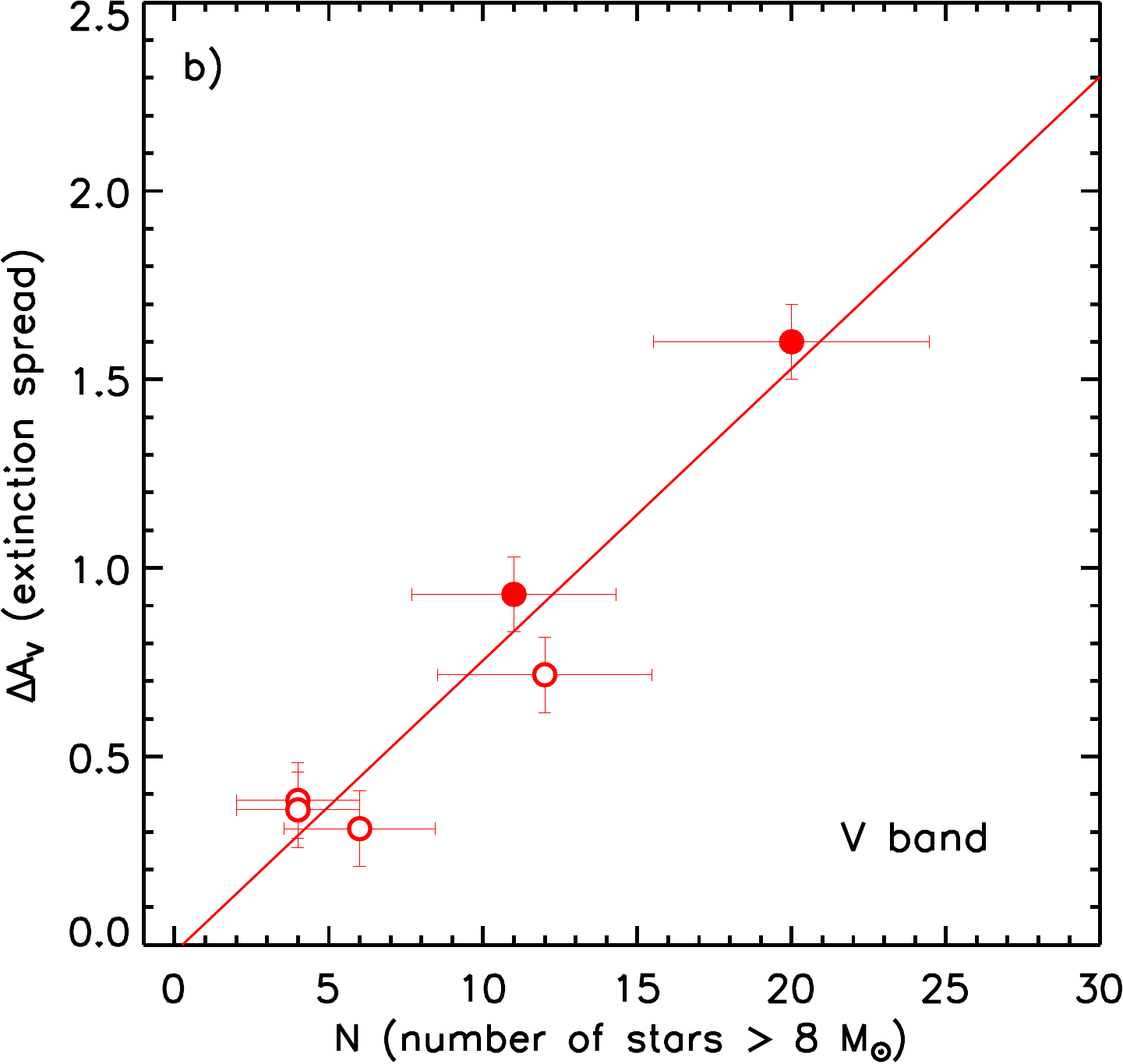}}
\resizebox{\hsize}{!}{\includegraphics{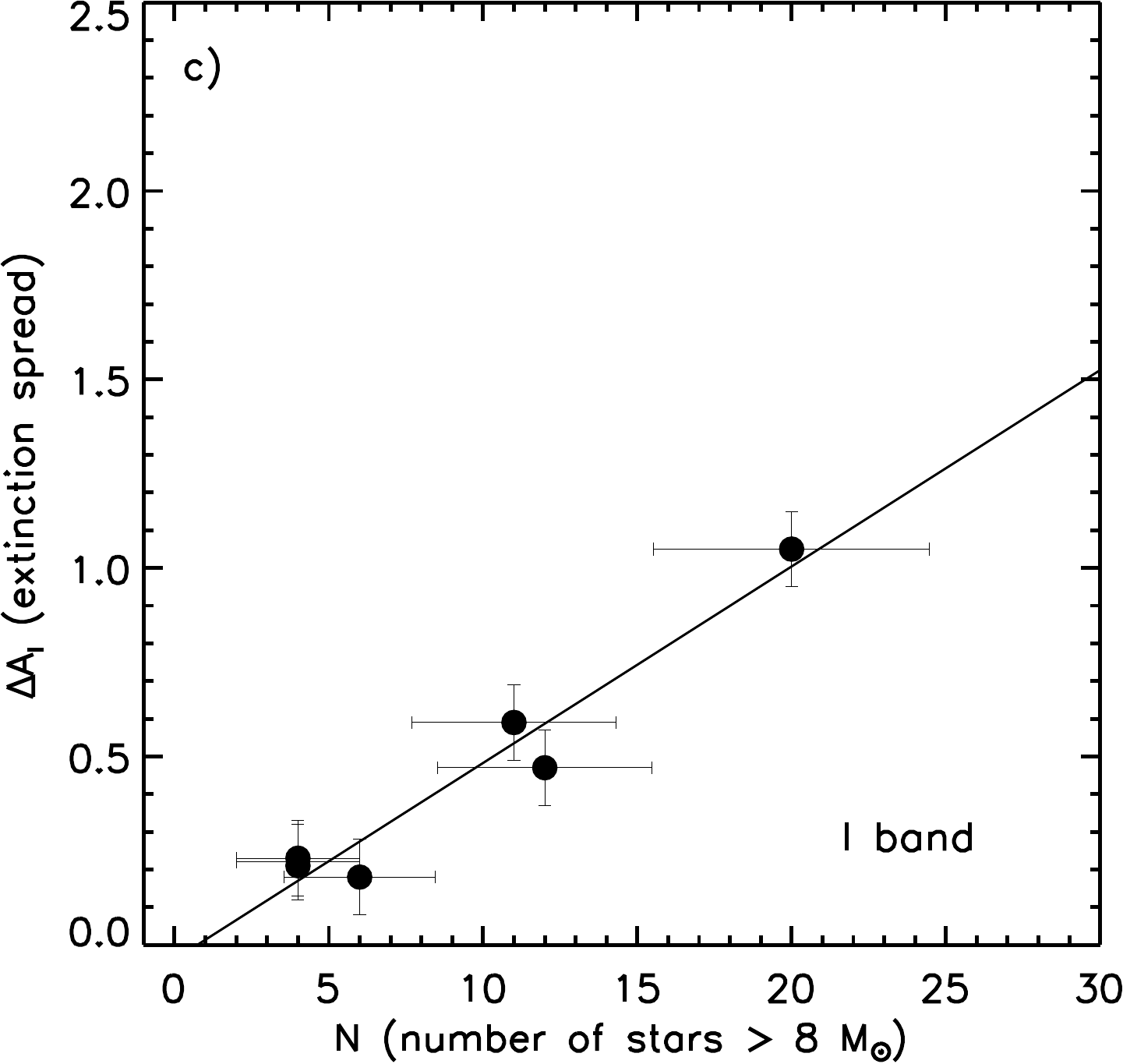}
                      \includegraphics{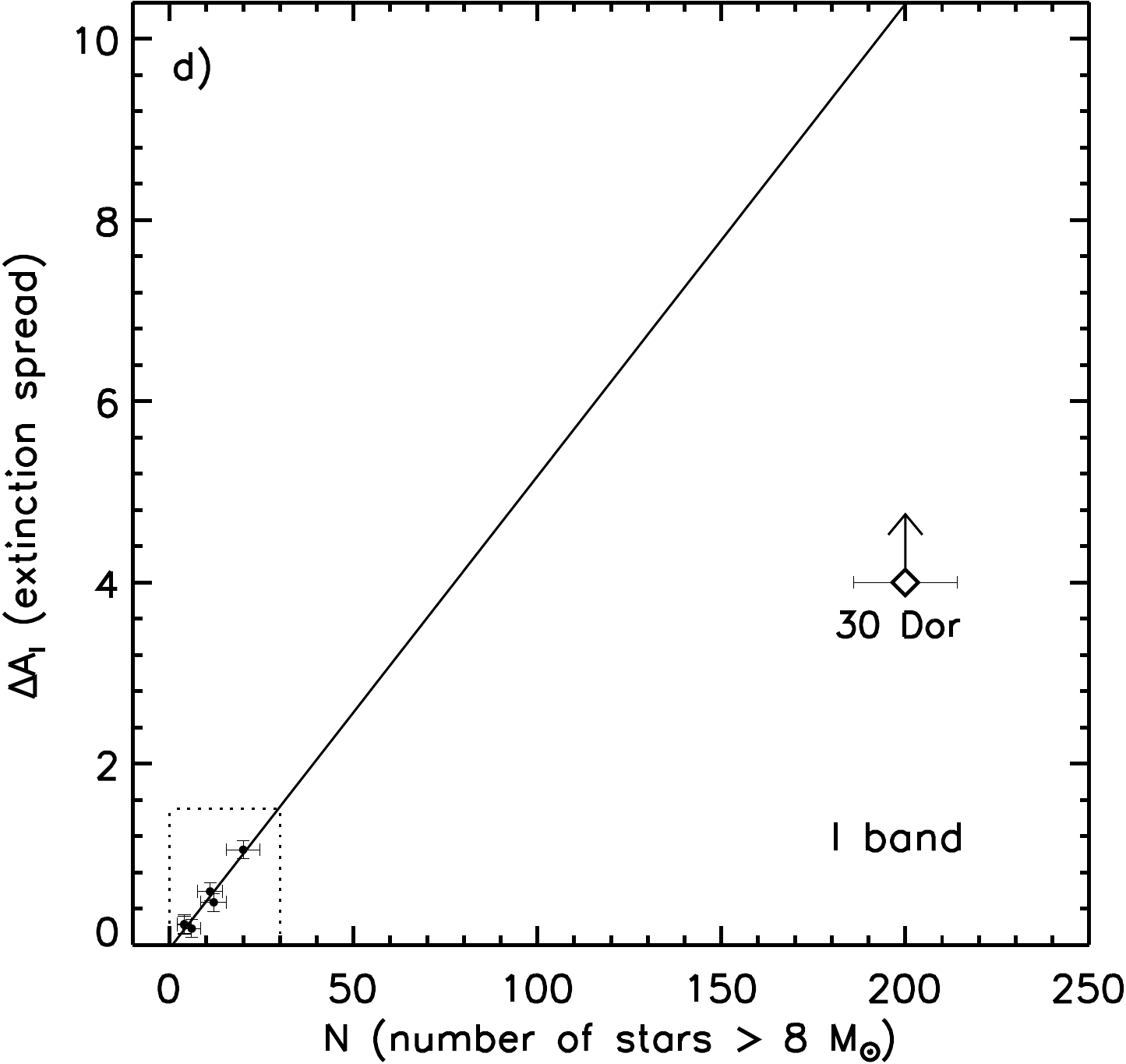}}
\caption{Extinction spread in the various regions, as a function of the
number of stars more massive than 8\,\Msolar. Measurements in the $B$,
$V$, and $I$ bands are indicated, respectively, in panels a), b), and
c). Filled dots mark extinction values measured directly in that
band, while empty dots are for measurements derived from neighbouring
bands. The lines represent the best linear fit in the three bands. Panel 
d) is equivalent to panel c), but the range is expanded to accommodate 
also the data point (lower limit) corresponding to 30\,Dor. }
\label{fig6}
\end{figure*}

Interestingly, the extinction range appears to increase from Field 1 to
Field 4 { around} NGC\,1856 and the growth continues further when
moving to NGC\,1854 and NGC\,1858.  A study of the young stellar
populations in these fields suggests that there is indeed a correlation
between the total amount of extinction and the number of massive stars,
as we show in the following.

To allow for a meaningful comparison between these fields, we counted in
each of them the number of massive stars with ages between 10 and 40
Myr, which in turn can be interpreted as a proxy for the recent
formation of  massive stars in these fields. This is achieved by first
building the CMDs from all stars present in each field, with no
distinction concerning their spatial location. We then drew on the CMDs
the isochrones extracted from the models of Chen et al. (2015) for our
specific bands and metallicity $Z=0.007$ and ages of 10 and 40 Myr, to
which we applied the same combination of foreground and intrinsic
extinction as in Figure\,\ref{fig2} -- \ref{fig4}. We finally counted
all stars with a MS mass of 8\,\Msolar or more, which were identified
with the help of the evolutionary tracks. The number of massive objects
selected in this way is shown in the last column of Table\,\ref{tab4}.
{ We note that the number of massive stars measured in this way is
necessarily subject to some uncertainty caused by reddening. Some of the
stars redder than the 40\,Myr isochrone might in fact be younger
objects subject to larger reddening. By the same token, objects younger
than 10\,Myr might appear redder than the corresponding isochrone. 
However, we expect the two effects to at least partly compensate,
thereby reducing the overall uncertainty. }

This analysis reveals that fields containing more massive stars also
have typically higher extinction and extinction spread. This is shown
graphically in Figure\,\ref{fig6}. The dots correspond to the values in
Table\,\ref{tab4} and the green, red, and black colours are used,
respectively, for the extinction in the $B$, $V$, and $I$ bands (Panels
a, b, and c; panel d, also referring to the $I$ band data, is discussed
further down). The uncertainties in the figure reflect the Poisson
statistics on the number of stars and a typical uncertainty of $0.1$ mag
on the measured extent of the RC, dictated by the intrinsic RC size
mentioned above. 

In the $I$ band we could directly measure $\Delta A_I$ in all regions
(black filled dots). In the $V$ and $B$ band, direct measurements are
available respectively for NGC\,1858 and NGC\,1854 (filled red dots),
and for { fields F1--F4 surrounding} NGC\,1856 (filled green dots).
Empty dots show values obtained from a neighbouring band by
interpolation through the extinction law { for} 30 Dor West (De
Marchi et al. 2014), since the latter is fully consistent with the
reddening vectors measured in all three clusters in the $B$, $V$, and
$I$ bands (see Section\,3).

All three sets of $\Delta A$ values are consistent with simple linear
correlations with the number of massive stars, as shown by the solid
lines. The formal coefficients are  $0.052$, $0.077$, and $0.093$ for
the three bands, respectively, with an uncertainty of about 13\%
($1\,\sigma$). Within the uncertainties, we cannot exclude that the
best-fitting lines have a zero intercept term. While certainly possible,
this does not appear to be likely, because it would imply that the
effects of star formation episodes on the local ISM vanish as soon as
the the last SNe have exploded, contrary to what is observed (e.g.,
Temim et al. 2015).

What is interesting, however, is not the actual values of the
coefficients, which necessarily depend on the selected sample and would
be different for another mass range, but rather the existence of a clear
correlation between extinction and the number of massive stars in these
fields. A linear correlation is also present if the mass range is
extended to stars down to 6\,\Msolar and the same age range
(10--40\,Myr). Further extending the study to less massive stars is not
possible without introducing large uncertainties on the age, since the
isochrones overlap. 

Selecting stars more massive than 8\,\Msolar is relevant because these
are the progenitors of core-collapse Type II supernovae (SNe II; e.g.
Edlridge \& Tout 2004) and we expect objects of this type to be likely
at the origin of the anomalous extinction properties that we see in
these and other LMC star-forming regions, as suggested by De Marchi \&
Panagia (2019) and De Marchi et al. (2020). Those works show that the
fraction and amount of big grains (with typical size $\sim 0.1\,\mu$m)
implied by the extinction laws measured in and around 30\,Dor and
NGC\,1938 are quantitatively consistent with the output caused by the
SNe II expected to have exploded in those regions in the past $\sim
50$\,Myr. The ejecta from these events appear quantitatively sufficient
(De Marchi et al. 2020) to have locally altered the standard grain-size
distribution typical of the diffuse ISM of the LMC (e.g., Gordon et al.
2003).

Figure\,\ref{fig6} suggests that one can estimate the range of the
extinction values in a region of star formation in the LMC by counting
the number of massive stars present in that field. The estimate is
necessarily approximate, and the relationships shown in
Figure\,\ref{fig6} are likely to become non linear and to saturate at
high extinction values, because highly extinguished RC stars cannot be
detected as such. 

An example is shown in Figure\,\ref{fig6}d, where the data from  panel
\ref{fig6}c (contained within the dotted rectangle) are compared with
the data point measured in 30 Dor. To obtain { the latter} point, we
used the photometric catalogue of De Marchi \& Panagia (2014), covering
the central $\sim 3 \times 3$ arcmin$^2$ of 30 Dor, together with the
median value of the extinction that they measured towards upper MS stars
in that field. We estimate that the central $\sim 3 \times 3$ arcmin$^2$
of 30\,Dor contain about 200 stars more massive that 8\,\Msolar and with
ages between 10 and 40 Myr. The most extinguished RC stars still
detectable as such in that field are about 4 mag fainter than the least
extinguished ones (see Figure\,6 in De Marchi \& Panagia 2014). However,
objects with even higher extinction (and $B-V>2$) might well be present
in the field, yet they are not detectable because of the limited depth
of the observations in the $B$ band. Therefore, the  corresponding data
point, $A_I\simeq 4$ (marked by a diamond in Figure\,\ref{fig6}d)
necessarily represents a lower limit to the maximum extinction value in
30\,Dor. That point appears to underestimate by about 60\% the amount of
extinction that one would obtain by extending the best fitting line in
Figure\,\ref{fig6}c to larger values of $N$. Even if there is no
guarantee that the relationship that we have measured in our smaller
clusters should remain linear in  star-forming regions as dense as 30
Dor, the solid line in Figure\,\ref{fig6}d is likely to provide a
realistic estimate of the maximum extinction to be expected in this
field, with an uncertainty of less than 50\%. In regions of less intense
star formation, we expect the uncertainty to be considerably smaller. 

\begin{figure*}
\centering
\resizebox{\hsize}{!}{\includegraphics{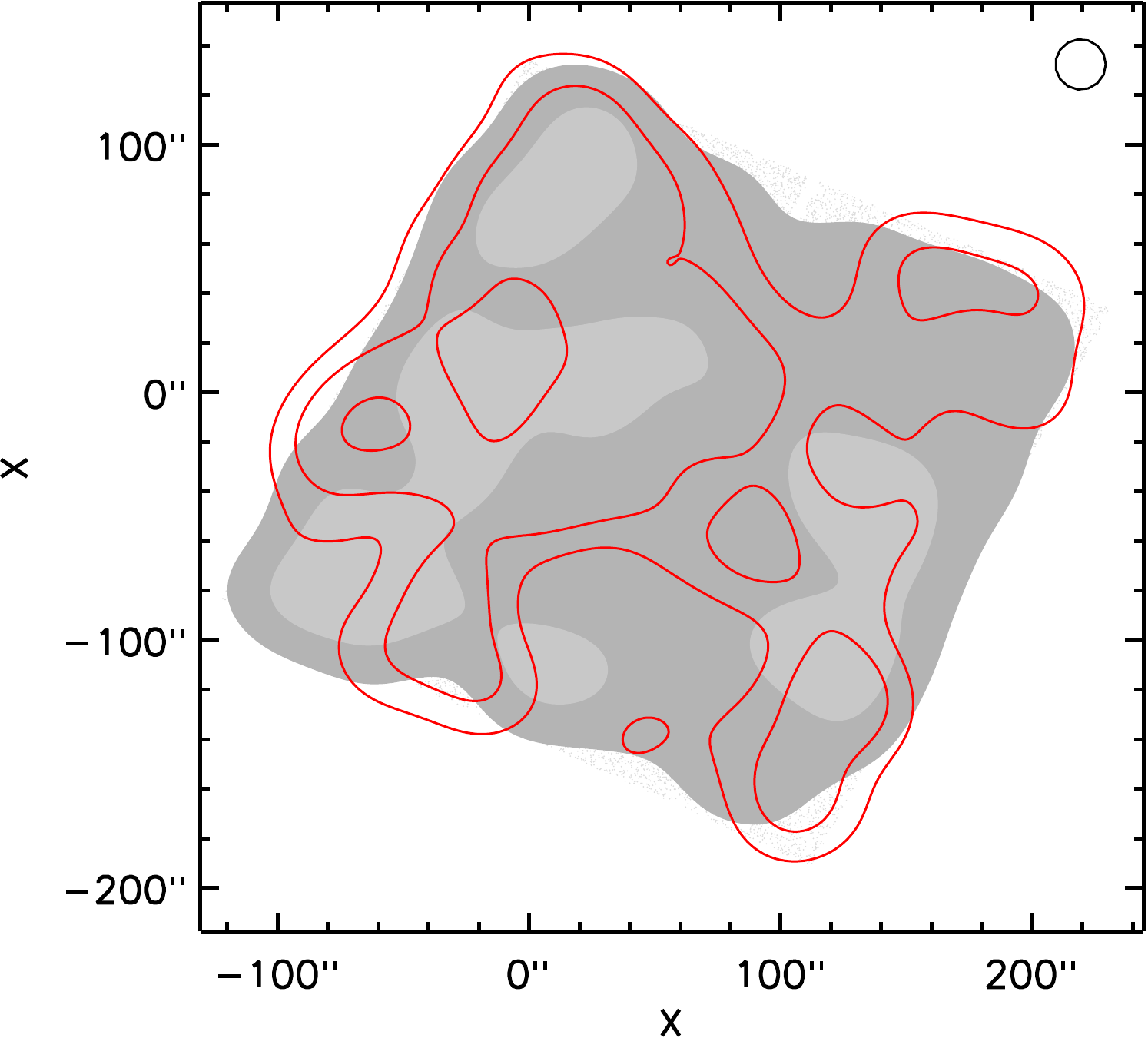}
                      \includegraphics{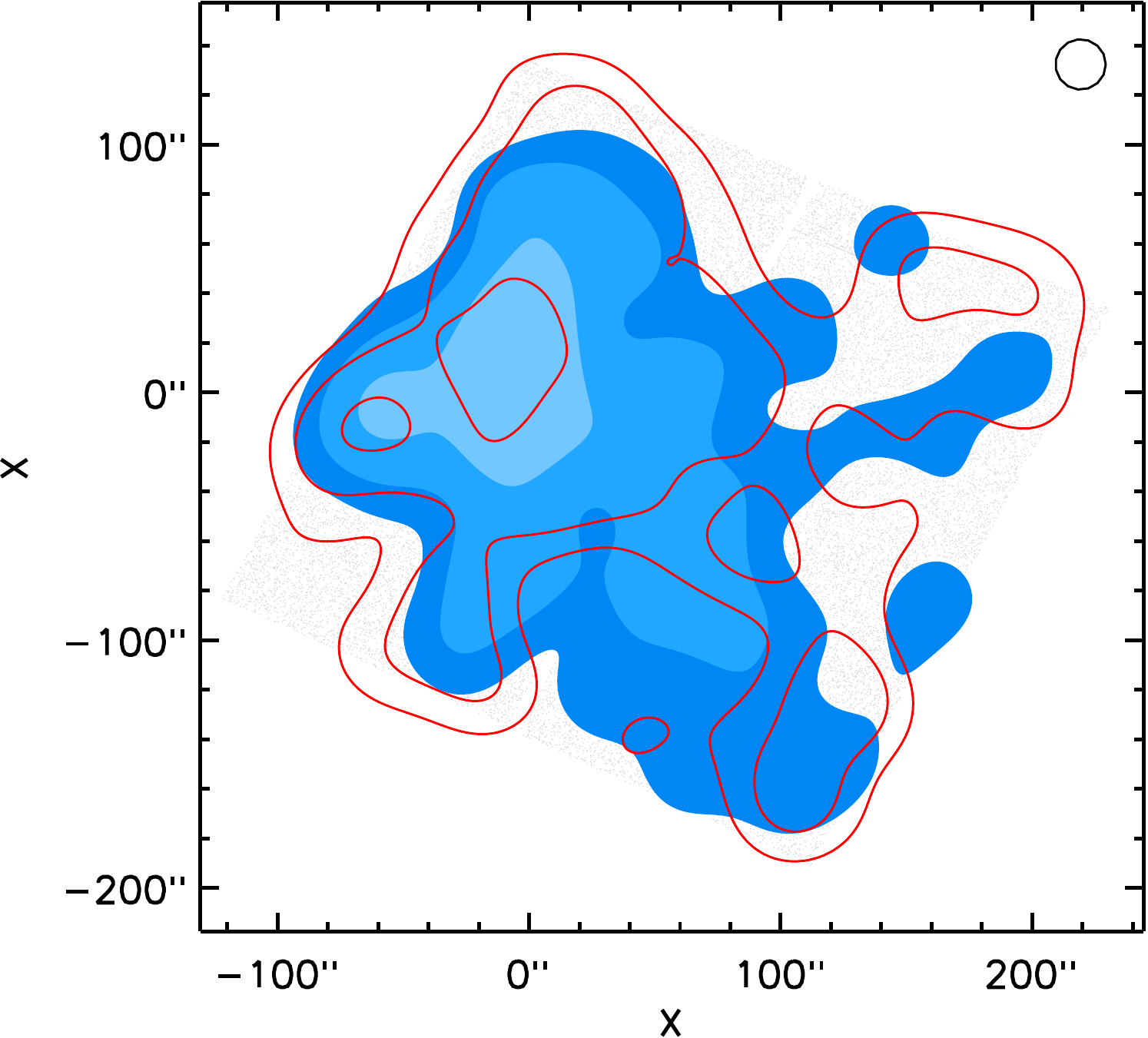}}
\caption{{ Contour plots showing the position and density
distribution of  RC and upper MS stars in the field containing NGC\,1858. The
coordinates system is centred on the nominal cluster centre. Red contour
lines are used for highly reddened RC stars ($A_V>1$), while the grey
and blue shadings are used for low-reddening RC stars and upper MS objects,
respectively. The lower contour level corresponds to twice the mean
density of each group. The circles at the top of the panels shows the
size of the Gaussian beam used for smoothing the actual distributions.}}
\label{fig7}
\end{figure*}

In summary, our analysis suggests that it is not the amount of
extinction that determines the extinction properties in a region, at
variance with what was generally concluded in the 1960s and 1970s from
the study of highly extinguished regions of the Galactic Plane (see
Introduction). Instead, it appears that recent massive-star formation is
systematically accompanied by ''non-standard'' grey extinction, whose
amount is correlated with the star formation strengtht.

{ Finally, to characterise the patchiness of the extinction in these
regions, we studied the spatial distribution of highly extinguished RC 
stars. We show an example in Figure\,\ref{fig7} for the case of the
region around NGC\,1858. We took as candidate RC stars all objects in
the CMD contained within $\pm 0.4$\,mag of the reddening vector shown 
in Figure\,\ref{fig5}a and fainter than $V=19$. A total of 582 stars
satisfy this condition: although some of them might in fact be RGB
stars, we expect the majority to be RC objects. About 80\,\% of them
have extinction $A_V<1.0$ and the remaining 20\,\% have values in the
range $1.0 < A_V < 1.5$. We compare the positions and spatial
distributions of these two groups in Figure\,\ref{fig7}a by means of
lines of stellar density with constant logarithmic steps set at 2, 4, and 8
times the mean density of each group. The contour plots  were obtained
after smoothing the actual distribution with a Gaussian beam with
$\sigma = 2\arcsec$, as indicated by the circle at the top of the
figures. The grey shaded contours correspond to the low-reddening RC
stars and the red contour lines to highly reddened RC objects.
Figure\,\ref{fig7}a reveals that the low-reddening RC stars are more
uniformly distributed than the RC objects with higher extinction. In
Figure\,\ref{fig7}b we compare, using the same relative contour levels,
the distribution of the highly reddened RC stars (red contour lines;
same as in panel a) with that of the massive  MS stars ($>8$\,\Msolar;
blue shaded contours). Although qualitative, this comparison shows that
the distribution of the highly reddened RC stars is rather similar to
that of the massive stars. This suggests that a considerable fraction of
the absorbing material along the lines of sight to these RC stars is
indeed associated with the region of recent star formation.}

\section{Conclusions}

We conclude with some considerations on the nature of the relationships
highlighted in Figure\,\ref{fig6}, which is not completely unexpected.
The typical dust extinction of star-forming galaxies is known to
increase with their total stellar mass $M_*$, not only in the local
universe but also out to redshift of at least $z \simeq 2$ (e.g.,
Brinchmann et al. 2004; Stasinska et al. 2004; Pannella et al. 2009). In
the range $10^8 \la M_*/$\Msolar$ \la 10^{10}$, the extinction has been
shown to increase monotonically and approximately linearly with $\log
M_*$ (Garn \& Best 2010; Zahid et al. 2013). Also the rate of star
formation and the metallicity of these galaxies correlate with the
extinction, but the predominant factor influencing it appears to be the
stellar mass (Garn \& Best 2010).

To be sure, the total mass of the star-forming regions probed by our
study is some $\sim 4$ orders of magnitude smaller than that of the
smallest of the high-redshift star-forming galaxies mentioned above. As
an example, adopting for NGC 1854 a standard initial mass function (e.g.
Kroupa 2001) with a power-law index $\gamma=-2.3$ for stars more massive
than $0.5$\,\Msolar and $\gamma=-1.3$ in the $0.08-0.5$\,\Msolar range,
we derived a total initial mass of about $2.5 \times 10^4$\,\Msolar for
this cluster. Nonetheless, the stars responsible for the bulk of the
dust, and particularly of the big grains that we detect in our regions,
are massive stars of the same type as those whose emission line fluxes
are used to infer both the total extinction and $M_*$ values of the
star-forming galaxies. These are the objects more massive than
8\,\Msolar that end their lives as SNe II (e.g. Dwek 1998). 

Although so far limited in size, our sample is potentially better suited
to studying the actual physical mechanisms at the origin of the observed
correlations. In our sample, the total mass, i.e. the number of stars,
is measured directly by counting individual objects, so we do not
have to rely on necessarily uncertain integrated quantities often
derived through approximate model-based relations (e.g. Kauffmann et
al. 2003): instead, we have direct measurements. Also the 
extinction properties (and extinction law) are directly measured by
us in our fields, including the value of $R_V$. Other studies instead
assume an average value of $R_V$, which is  known to be affected by
large uncertanties (Calzetti et al. 2000) and is not applicable outside
the regime of centrally-concentrated starburst galaxies (Calzetti et al.
2021). Furthermore, the fact that the empirical relationships
that we discovered from a limited number of small star-forming regions
is able to reproduce, to within a factor of two, also to the case of 30
Dor suggests that these relationships are meaningful also for the
intense star-forming knots of high-redshift galaxies that 30 Dor mimics
(e.g., Doran et al. 2013; Crowther et al. 2017). Extending our study of
the extinction properties to a number of other star-forming regions in
the LMC (De Marchi et al., in prep) will allow us to better constrain
the nature of the physical processes at play.

\vspace*{0.5cm}

We are very grateful to an anonymous referee whose expert advice and
costructive criticism have helped us to improve the presentation of this
work. APM acknowledges support from the European Research Council (ERC)
under the European Union's Horizon 2020 research innovation programme
(Grant Agreement ERC-StG 2016, No 716082 'GALFOR',
http://progetti.dfa.unipd.it/GALFOR), by the MIUR through the FARE
project R164RM93XW 'SEMPLICE' and the PRIN programme 2017Z2HSMF.

\end{document}